\documentclass[12pt,preprint]{aastex}

\slugcomment{Accepted by The Astronomical Journal}

\shorttitle{Water masers in Bok globules}
\shortauthors{G\'omez et al.}

\begin{document}

\title{A sensitive survey for water maser emission towards Bok
  globules using the Robledo 70m antenna}

\author{Jos\'e F. G\'omez\altaffilmark{1}, Itziar de
  Gregorio-Monsalvo\altaffilmark{2}, Olga Su\'arez\altaffilmark{2},
  Thomas B. H. Kuiper\altaffilmark{3}}

\altaffiltext{1}{Instituto de Astrof\'{\i}sica de Andaluc\'{\i}a,
  CSIC, Apartado 3004, E-18080 Granada, Spain; e-mail: jfg@iaa.es}
\altaffiltext{2}{Laboratorio de Astrof\'{\i}sica Espacial y F\'{\i}sica
  Fundamental, INTA, Apartado 50727, E-28080
    Madrid, Spain; e-mail: itziar@laeff.esa.es;
    olga@laeff.esa.es} 
\altaffiltext{3}{Jet Propulsion Laboratory, California Institute of
  Technology,  4800 Oak Grove Dr, Pasadena, CA 91109 USA}

\begin{abstract}
We report the most sensitive water maser survey towards Bok globules
to date, using NASA's 70m antenna in Robledo de Chavela (Spain). We
observed 207 positions within the \citet{cb} catalog with a higher
probability of harboring a young star, using as selection criteria the
presence of radio continuum emission (from submillimeter to centimeter
wavelengths), geometrical centers of molecular outflows, peaks in maps
of high-density gas tracers (NH$_3$ or CS), and IRAS point sources. We
have obtained 7 maser detections, 6 of which (in CB 34, CB 54, CB 65,
CB 101, CB 199, and CB 232) are reported for the first time here. Most
of the water masers we detected are likely to be associated with young
stellar objects (YSOs), except for CB 101 (probably an evolved object)
and CB 65 (uncertain nature). The water maser in CB 199 shows a
relatively high shift ($\simeq 30$ km s$^{-1}$) of its velocity
centroid with respect to the cloud velocity, which is unusual for
low-mass YSOs. We speculate that high-velocity masers in this kind of
object could be related with episodes of energetic mass-loss in close
binaries. Alternatively, the maser in CB 199 could be pumped by a
protoplanetary or a young planetary nebula. CB 232 is the smallest Bok
globule ($\simeq 0.6$ pc) known to be associated with water maser
emission, although it would be superseded by the cases of CB 65
($\simeq 0.3$ pc) and CB 199 ($\simeq 0.5$ pc) if their association with YSOs
is confirmed. All our selection criteria have statistically compatible
detection rates, except for IRAS sources, which tend to be a somewhat
worse predictor for the presence of maser emission.
\end{abstract}

\section{Introduction}

Bok globules \citep*{bok47} 
are small (projected size $\la 20'$), isolated, and
relatively simple  molecular
clouds, with typical masses of $\simeq 5-50$ M$_\odot$
\citep*{mar78, cle91}.
These globules are usually identified and cataloged 
as dark patches in optical images (e.g., \citealt*{cb}, hereafter CB). 
The reference catalogs of Bok 
globules compiled by CB and  \citet*[hereafter BHR]{bhr}
in the north and south
hemispheres, respectively, focused on the group of smaller Bok globules, using a size of
$<10'$ as a selection
criterion. Under the assumption that
they are nearby clouds ($d\la 500$ pc), as suggested by the small
number of foreground stars, the globules in the CB and BHR catalogs
would in general have linear sizes $\la 1$ pc.

Some Bok globules are sites of low- and
intermediate-mass
star  formation (e.g., \citealt*{yun91}, \citeyear{out-yc92}; \citealt*{rei92}).
Since they are identified from optical
images, without any characterization of their possible star-forming
activity, catalogs of Bok globules may span a wide range of
evolutionary stages, from quiescent dark cores to clusters of Herbig
Ae/Be or T-Tauri stars. Therefore, systematic studies of these
globules are potentially useful to study the evolution of phenomena
related to  star formation, like collapse, fragmentation, mass-loss,
or formation of protoplanetary disks. Moreover, being small and relatively
simple molecular clouds, one can study these phenomena
with a lower
chance of contamination from multiple generations of young stellar
objects (YSOs) within
the same region. For these reasons, they are interesting laboratories in
the study of the star-formation processes and its evolution.

Another important reason to study star formation in Bok globules is
that field stars in the solar neighborhood, and the Sun itself, may have
formed in this kind of cloud. Moreover, some T Tauri stars
apparently not related to known molecular clouds, may have originated
in Bok globules which have dispersed, leaving isolated 
pre-main-sequence objects \citep*{mm-lh97}.

An important phenomenon related to star formation is the occurrence of 
water maser emission at 22 GHz. 
This emission is a powerful tool to study the morphology and kinematics
of the environment of YSOs at high angular and spectral resolution
\citep*{tor02,cla05}.
The physical conditions needed for maser excitation can be reached in
both the inner parts of circumstellar disks surrounding the YSOs and in
shocked gas related to winds. This  dichotomy has been
suggested to be an evolutionary effect by \citet*{tor97,tor98}, 
in which masers may trace bound motions (e.g.,
disks) in the
youngest protostars, and  outflows in  more evolved
objects.   

Although water masers are more intense and widespread in high-mass
star forming regions, they are also present around low-mass YSOs
\citep*{wil87,ter92}. In the case of low-mass star formation, there
seems to be an 
evolutionary trend, with water masers tracing the youngest YSOs,
i.e., Class 0 protostars. In their survey of water maser emission
towards low-mass YSOs, \citet*{fur01} obtained a  detection rate of
40$\%$ on Class 0 sources, 4$\%$ on Class I, and 0$\%$ on Class II.

Since there are several evolutionary aspects related to water maser 
emission, it is interesting to study this emission in Bok globules,
where we can find YSOs in different evolutionary stages. 
However, there are very
few studies of water masers towards Bok globules. 
Only \citet*{sca91} conducted a search for water masers
specifically in Bok globules. These authors observed 80 globules, and obtained
a detection towards CB 3 (LBN 594), with a peak flux density of $\simeq 77.6$
Jy. However, their detection threshold ($\sim 4.4-15.7$ Jy) would miss
a high fraction of masers around low-mass YSOs.

Other surveys included among their
targets some objects within the globules in the CB catalog
\citep*{fel92,pal93,wou93,per94,cod95}. To our knowledge, only three Bok
globules of the CB catalog have been reported to harbor a water
maser: the mentioned one in CB 3 \citep{sca91}, a detection of $\simeq
1-2$ Jy towards the Herbig Be
star HD 250550 in CB 39 \citep*{sch75}, and another one of $\simeq
7.5$ Jy
in CB 205 (LDN 810) obtained by \citet*{nec85}. We are not aware of any
water masers in the southern Bok globules of the BHR catalog. However, all these three
globules (CB 3, CB 39, and CB 205) are somewhat different from most of the
clouds in the CB catalog, since they are located at distances $\geq 1$ kpc,
and thus are larger and more massive than the globules that
CB intended to compile.

In this paper, we present the most extensive and sensitive survey of
water masers in Bok globules to date. We have searched for this emission
towards the positions within the globules in the CB
catalog with the highest probability of harboring a YSO (see Sec. \ref{sec:sample}). This survey
aims to be nearly complete for possible star-forming regions in Bok
globules north of declination $\simeq -36^\circ$, to study the conditions and characteristics of 
maser emission
in these isolated globules. This work is complemented by 
high-resolution interferometric studies of the detections in this
survey, to accurately determine their spatial and velocity
distribution, and to pinpoint their exciting source  (de
Gregorio-Monsalvo et al. 2006, hereafter DG06). 
The present paper is structured as follows: In
Sec. \ref{sec:observations} we describe the technical details of the
observations, in Sec. \ref{sec:sample} we explain the criteria used to
select the target sources, and we present our results in
Sec. \ref{sec:results}, which are further discussed in
Sec. \ref{sec:discussion}. We summarize our conclusions in
Sec. \ref{sec:conclusions}.

\section{Observations}
\label{sec:observations}
We observed the  6$_{16} \rightarrow$5$_{23}$ transition of the water molecule 
(rest frequency 22235.080 MHz),
using
the NASA 70-m antenna (DSS-63) at Robledo de Chavela (Spain).
The 1.3 cm receiver of this antenna was a cooled
high-electron-mobility transistor (HEMT). 
Water maser observations were carried out using three different
backends depending on the observing dates: From 2002 March 13 to 2002 April 10, we used a 4096-channel spectrometer covering a 
bandwidth of 400 MHz ($\simeq 5398$ km s$^{-1}$), which provided a velocity
resolution of $\simeq 1.3$  km
s$^{-1}$. From 2002 April 14 to 2003 July 18, we used a 
256-channel spectrometer, covering a bandwidth of 10
MHz, which provided a velocity
resolution of $\simeq 0.5$ km s$^{-1}$. From 2004 July 6 to 2005
October 16, we used a 384-channel
spectrometer, covering a bandwidth of 16 MHz ($\simeq 216$ km s$^{-1}$
with $\simeq 0.6$  km s$^{-1}$
resolution). 
At this frequency, the half-power beamwidth
of the telescope is $\simeq 41''$. Spectra were taken in
position-switching mode with the 4096- and 384-channel spectrometers,
and in 
frequency switching mode, with a switch of 5 MHz, when using the
256-channel one, thus providing in the latter case an
effective
velocity coverage of $\simeq 202$ km s$^{-1}$ (15 MHz) centered at the $V_{\rm 
LSR}$ of each source. Only left circular polarization was processed.
System temperatures ranged between 45 and 135 K, and the total integration
time was typically 20 min per source in frequency-switching mode and
30 min (on+off) in position-switching mode. The rms pointing accuracy was
better than 10$''$. The data reduction was performed using
the CLASS package, which is part of the GILDAS software. 

\section{Source Sample. Selection criteria}

\label{sec:sample}
The target
positions in our survey 
are those within the Bok globules cataloged by CB that show indications of possible star formation, or with
higher probability of harboring a YSO. We used four selection criteria for
those targets: 
\begin{enumerate}
\item Radio continuum sources (submillimeter to centimeter wavelengths).\\
  Low-mass stars undergoing mass-loss can show radio continuum
  emission from ionized winds \citep*{cm-a92}, 
  which at cm wavelenghts is not likely
  to be contaminated by dust emission from envelopes and disks. For
  the mm and submm emission, the youngest protostars (deeply embedded
  class 0 sources) show prominent emission at those wavelengths
  \citep*{and93,sar96}. The 
  radio continuum sources were taken mainly from
  \citet*{cm-a92, cm-a98}, \citet*{cm-y96}, and \citet*{cm-m99} at cm
  wavelenths, \citet*{mm-lh97} at mm, and
  \citet*{smm-l97}, \citet*{smm-h99}, \citet*{smm-h00}, and \citet*{smm-v02} at
  submm ones. We selected
  those sources located  within the optical boundaries of the
  cloud and which were not suspected to be extragalactic objects due
  to their negative spectral indices \citep[see e.g.,][]{cm-a98}.
\item Center of
molecular outflows.\\
Molecular outflows driven by low-mass YSOs are specially
powerful during the earlier evolutionary phases \citep*{bon96},
which in their turn, are more likely to show water maser emission
\citep*{fur01}. We
have chosen the position of the geometrical center of the outflow
(i.e. between the red- and blue
shifted lobes), since we expect the
driving source of the outflow to be near that position. In the case of
outflows with only one lobe, we selected the position of the proposed
powering source. Most molecular
outflows known in Bok globules of the CB catalog 
have been reported by \citet*{out-yc92,out-yc94}.
\item Peak of high-density molecular gas tracers.\\
YSOs form in the densest part of molecular clouds. The maxima in maps
of high-density gas tracers like NH$_3$ have been found to pinpoint the
location of the driving sources of molecular outflows
\citep*{ang89,gom94}. 
For this survey
we have used the peaks of NH$_3$ maps compiled by \citet*{jij99}, most
of them observed by \citet{lem96} in these globules, and peaks of CS
maps published by \citet*{cs-l98}.
\item IRAS sources.\\
Many globules have been scarcely studied, if at all, so the previous
observational signs of possible star formation are not always
available. Therefore, 
we have also used the
position of IRAS sources as possible locations of YSOs. 
We considered as target positions all IRAS sources listed by CB as
associated with the globules in their catalog. Most of these IRAS
sources show rising IRAS fluxes at longer wavelenghts, which is
expected for  class 0 and class I objects  \citep{wil89,and93}, but
our search was not restricted based on their far infrared spectra.
\end{enumerate}

These criteria are ordered by decreasing preference given to the observations
of sources selected by each of them.
After compiling  the initial set of target positions (all located
north of $\delta = -36^\circ$)
we dropped from the list
most sources lying within a distance of $\sim 21''$ from 
another source with equal or
higher preference, since they would be within the beam of Robledo
radio telescope, although we observed some positions that would
have been left out by this proximity filter, specially for the ones
that would fall closer to the edge of the telescope beam.
With these selection criteria, the previously reported masers in CB 39
and CB 205
are within the beam of one of the positions in our survey (see notes
to Table \ref{tbobserved}). This is not
the case for the maser reported in CB 3 by \citep{sca91}, but DG06 found a more accurate
position with interferometric data, which lies $5''$ from
the position of CB3-mm, a source that is included in our target
list. Therefore, we see that our selection criteria seem appropriate
to locate water maser emission that may exist in Bok globules of
the CB catalog.

Using the final list of target sources, we searched for water masers
around all 
radio-continuum sources, centers of outflows, and peak of high-density
tracers, with a total of 100 observed positions. 
In addition to that, we also observed 107 IRAS
sources with no known nearby source falling into the other three
categories. The observed sources are shown in
Table \ref{tbobserved}. Note that our survey includes, in many cases, more
than one target position within the same globule. A total of 103 globules
were covered by our observations.

Considering the sources lying within a distance of $\sim 21''$ from
each observed position, and therefore, which fell within the telescope
beam for 
at least one of the observations, our survey covered 34 centimeter
sources, 20 millimeter sources, 30
submillimeter sources, 16 centers of outflows, 16 peaks of CS maps, 18
peaks of NH$_3$ maps, and 132 IRAS sources.

\section{Results}
\label{sec:results}
\subsection{Survey results}

\label{sec:surveyresults}
\begin{figure}
\rotatebox{-90}{
\epsscale{0.5}
\plotone{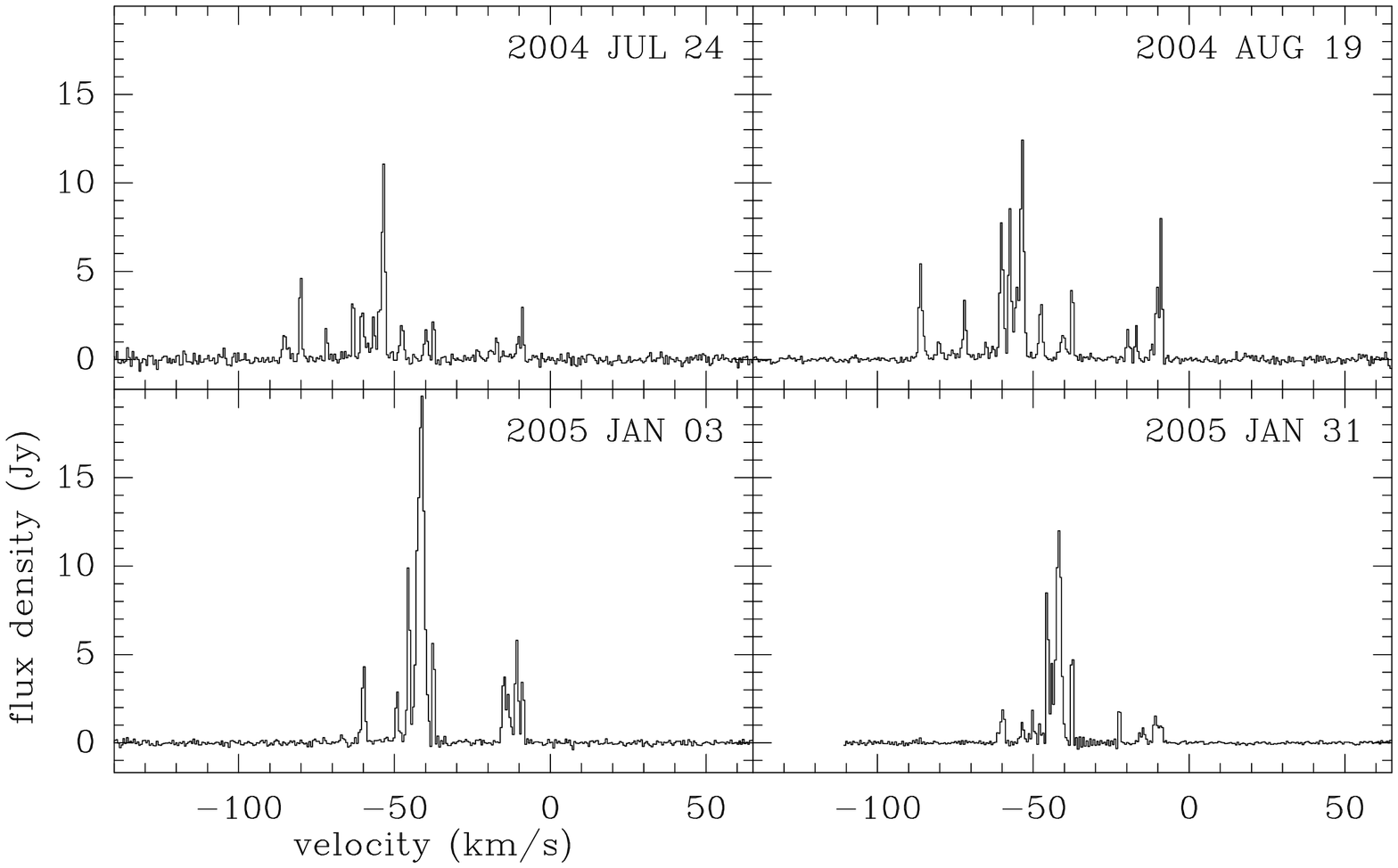}}
\caption{Water maser spectra towards CB3-mm}\label{fig:cb3}
\end{figure}

\begin{figure}
\rotatebox{-90}{
\epsscale{0.30}
\plotone{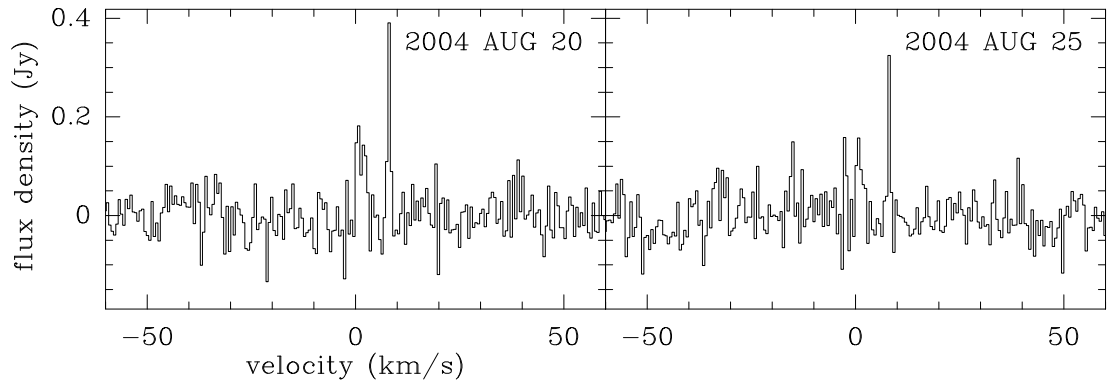}}
\caption{Water maser spectra towards [HSW99] CB 34 SMM 3}\label{fig:cb34}
\end{figure}

\begin{figure}
\rotatebox{-90}{
\epsscale{0.7}
\plotone{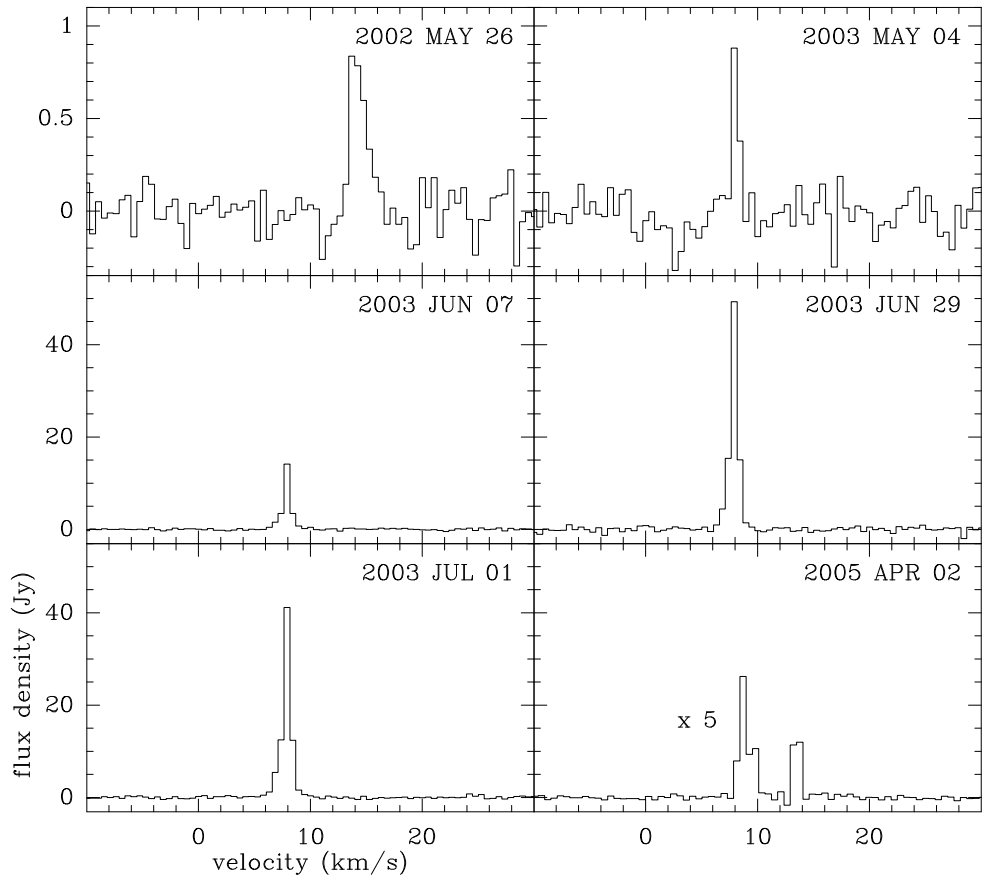}}
\caption{Water maser spectra towards [YMT96] CB 54 2. Note that the
  scale of the top panels is different from the rest. The spectrum
  taken on 2005 April 02 (bottom right) has been multiplied by a
  factor of 5 to show it more clearly}\label{fig:cb54}
\end{figure}

\begin{figure}
\rotatebox{-90}{
\epsscale{0.5}
\plotone{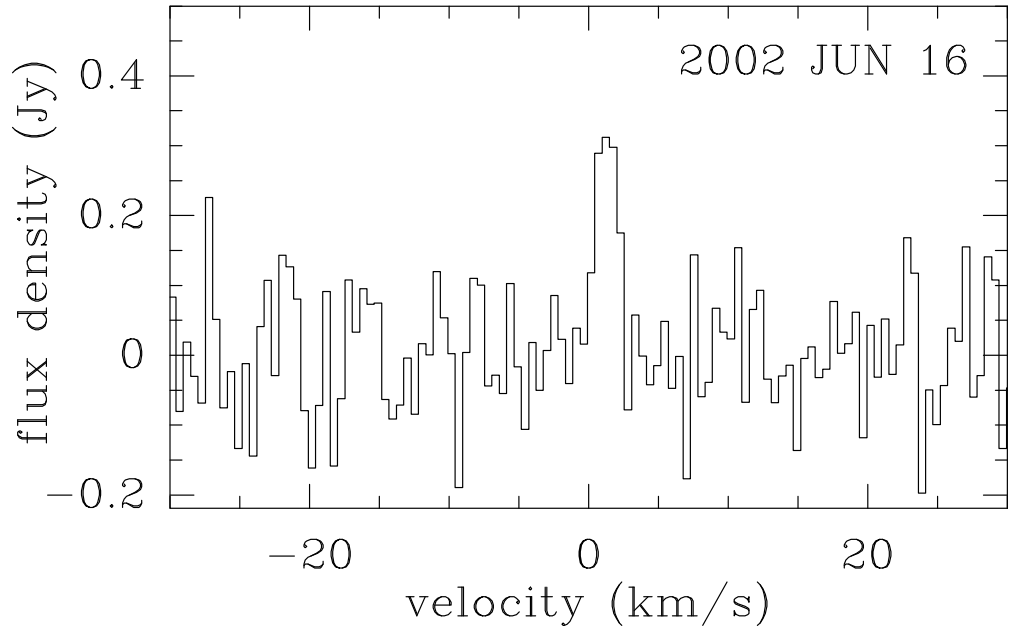}}
\caption{Water maser spectrum towards IRAS 16277-2332 in CB 65}\label{fig:cb65}
\end{figure}

\begin{figure}
\rotatebox{-90}{
\epsscale{0.7}
\plotone{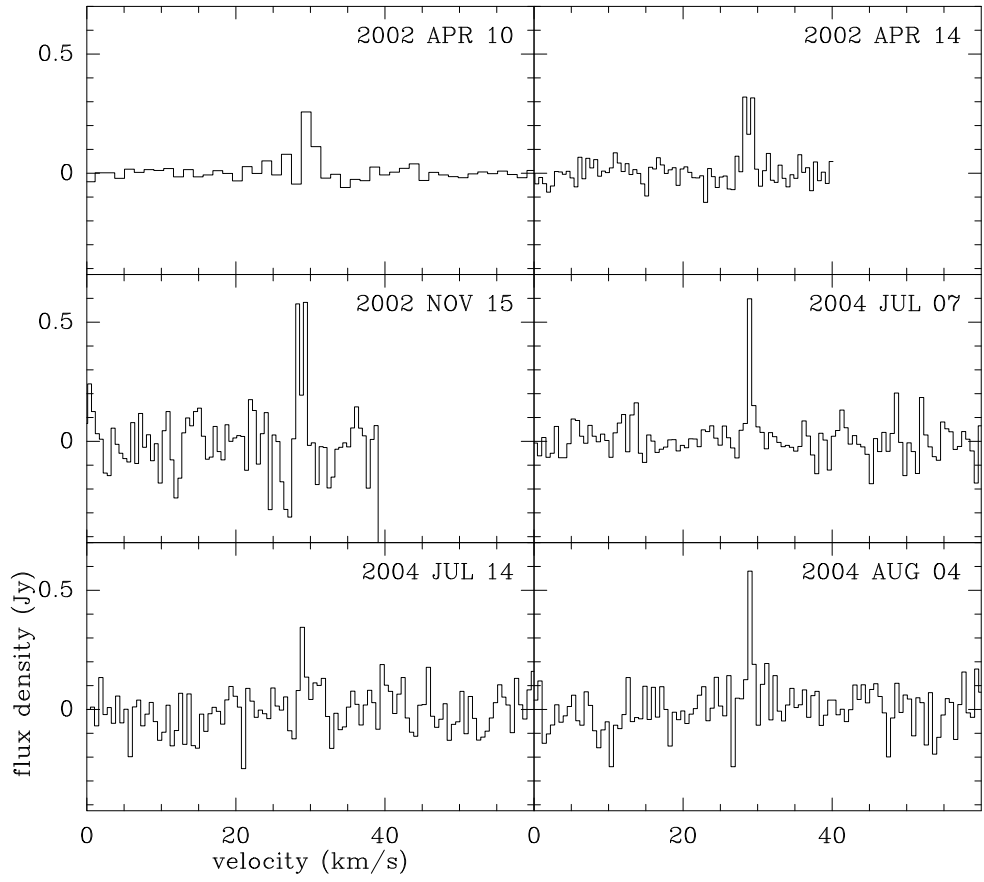}}
\caption{Water maser spectra towards IRAS 17503-0833 in CB 101}\label{fig:cb101}
\end{figure}

\begin{figure}
\rotatebox{-90}{
\epsscale{0.3}
\plotone{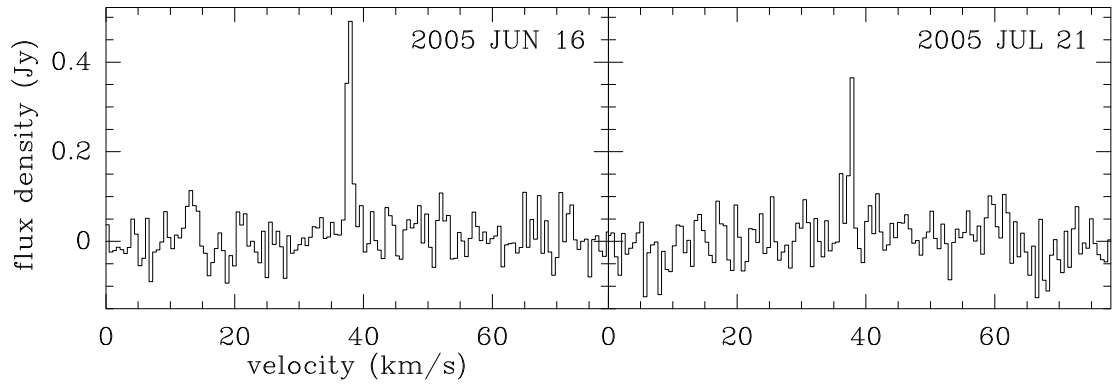}}
\caption{Water maser spectra towards [ARC2001] HH 119 VLA 3 in CB
  199}\label{fig:cb199} 
\end{figure}

\begin{figure}
\rotatebox{-90}{
\epsscale{0.5}
\plotone{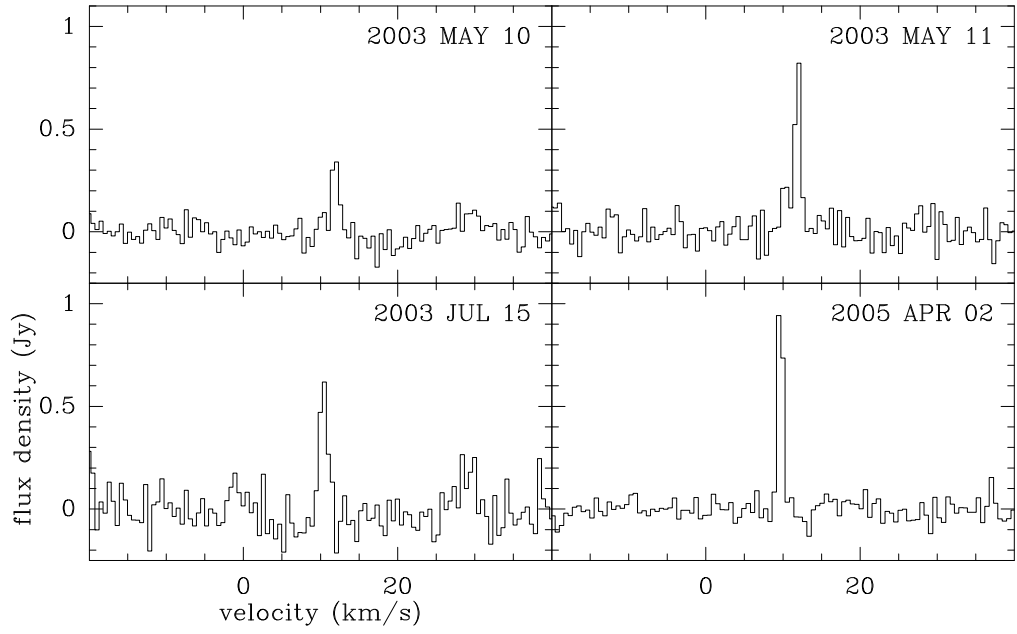}}
\caption{Water maser spectra towards IRAS 21352+4307 in CB 232}\label{fig:cb232}
\end{figure}

Tables \ref{tbdetections} and \ref{tbnondetections}
show the result of our survey of water maser
emission in Bok globules. We have obtained 7 detections
(Table \ref{tbdetections}), 6 of which (the ones in 
CB 34, CB 54, CB 65, CB 101, CB 199,
and CB 232) are reported for the
first time here. Of the previously known masers, only CB3-mm was detected
by us. We did not detect the masers in either CB 39 or CB 205. 
In Figs. \ref{fig:cb3} to \ref{fig:cb232} we show the spectra of the detected masers.

With these detections, we have increased the number of sources  in
Bok globules known to emit water maser emission from 3 to 9. However,
we note that some of the new detections are not likely to be related to their
respective globules, and they could be background or foreground
objects projected against the clouds (see Sec. \ref{sec:individual}). The ones which are
most likely to be associated with YSOs in these globules are the masers
in CB 34, CB 54, and CB 232 (and probably CB 199).

With respect to our selection criteria, considering the sources that
fall within the beam of the telescope of a position with water maser
detected, we have detected water maser emission around 2 centimeter
sources (6\%), 2 millimeter sources (10\%), 4 submillimeter sources 
(14\%), 3 centers of
molecular outflows (19\%), 3 peaks of CS (17\%),
no peak of NH$_3$ (0\%), and 5 IRAS sources (4\%), 
2 of them without association with any of the other
criteria (2\%). 

\subsection{Detections: individual sources}
\label{sec:individual}
\subsubsection{CB 3 (CB3-mm)}

CB 3 is located at a distance of
$\simeq 2.5$ kpc \citep*{mm-lh97}. This large distance (and thus, large
physical size, $\simeq 4.5$ pc, see CB), 
and the presence of
 intermediate-mass star
formation (it hosts a source of $L_{\rm bol}\simeq 930$ M$_\odot$; \citealt{smm-l97})
distinguish it from the rest of the globules in the CB catalog.

We detected water maser emission towards the millimeter source CB3-mm
detected by \citet{mm-lh97}. 
This source is believed to be the powering source of
a powerful bipolar molecular outflow \citep*{out-yc92,out-yc94, cod99}. The
maser probably 
corresponds to the one detected by \citet{sca91}, although they reported  a
position $\sim 1'$ from CB3-mm. We took spectra towards their position, but
we did not detect any emission ($3\sigma$ upper limit of $0.5$ Jy on
2004 August 19, between $V_{\rm LSR} = -146.2$ and 69.5 km s$^{-1}$), 
which is compatible with beam response
if the emission is actually related with CB3-mm. 
Within the beam of the
Robledo telescope from CB3-mm, 
there is also submillimeter emission \citep{smm-l97, smm-h00}, the peak of a
CS map \citep*{cs-l98}, and the source IRAS 00259+5625
(Table \ref{tbobserved}).

The maser emission is relatively strong ($10-20$ Jy), and shows variations 
of a factor of
$\sim 2$ (Table 3). The spectra are very rich in maser components at
different velocities (Fig. \ref{fig:cb3}), with variations in the velocity
distribution and in the ratio between different components. The data 
taken in 2004 show maser emission over a
wider range of velocities (from $-85$ to $-10$ 
km s$^{-1}$), while
those taken in 2005 do not show emission more blueshifted than $-60$
km s$^{-1}$. The centroid velocity of the maser emission 
($V_{\rm LSR}= -37.5$ to $-52.8$ km s$^{-1}$, depending on the observing
date) is
within 15  km s$^{-1}$ from the cloud
velocity ($V_{\rm LSR}\simeq -38.3$  km s$^{-1}$, CB).

The geometry of H$_2$ knots suggest that CB3-mm is
ejecting a precessing jet \citep*{mas04}. Interferometric
observations of the water maser emission are also compatible with this
suggestion (DG06).

Given the large number of different velocity components, and their
time variation, it would be interesting to carry out 
an interferometric monitoring of this source, to trace the spatial
distribution and proper motions of these different components, to
test the proposed scenario of a precessing jet, and to
ascertain whether the maser variability is related with the presence 
of episodic mass-loss phenomena.

\subsubsection{CB 34 ([HSW99] CB 34 SMM 3/SMM 4)}

CB 34 is a globule with multiple star formation \citep*{alv95}, located at 1.5
kpc \citep*{mm-lh97}. 
We detected water maser emission towards the submillimeter source
CB 34 SMM 3 \citep{smm-h00}. 
The maser emission was detected in August 2003, with two main
components (at $\simeq 1$ and 8 km s$^{-1}$, respectively) 
close to the cloud velocity ($V_{\rm LSR}\simeq 0.7$ km s$^{-1}$, CB). 
Several months
later, the emission dropped below the detection threshold of the telescope. 

In the vicinity of CB 34 SMM 3
there are several Herbig-Haro objects (HH 290N1, HH 290N2,
HH290S, and HH291) and H$_2$
knots, all forming at least three highly collimated jets in different
directions  \citep*{mor95, kha02}. Within the Robledo beam from CB 34
SMM 3 also lies the submillimeter source CB 34 SMM 4 \citep{smm-h00}, and the
geometrical center of
the bipolar outflow reported by \citet*{out-yc92}, oriented NE-SW, although
higher-resolution observations \citep*{kha02} seem to locate its
center closer to the
position of CB 34 SMM 1. Although CB 34 SMM 4 was not in our original
list of targets, given that it was within the beam when observing at 
CB 34 SMM 3, we took an additional spectrum at its position
[$\alpha(J2000)= 05^h 47^m 05\fs 2$, $\delta(J2000) = +21^\circ 00' 25''$] on 2004
August 25, to try to determine with which source the maser emission
was more likely to be related to. However, the spectrum was similar (peak
intensity $\simeq 0.32\pm 0.13$ mJy at 7.7 km s$^{-1}$) to the one obtained at
CB 34 SMM 3 the same day (Table \ref{tbdetections}). 
Therefore, it is not possible to ascertain
the association of
the maser with either submillimeter source, and since the
signal-to-noise ratio of the spectra was relatively low, any other
attempt to determine a more precise position would have been subject to high
uncertainties. The water maser may in fact be pumped by a
source lying roughly at the same distance from SMM 3 and SMM 4. A good candidate
could be source Q, also lying within the telescope beam from both
submillimeter sources (at $\sim 13''$ from them), 
and which \citet*{mor95} suggested to be the powering
source of one of the jet-like chains of H$_2$ knots (Q knots). Interestingly,
this jet is the one whose orientation is closer to that of the
molecular outflow. Therefore, this Q-jet could trace the dominant
mass-loss process in CB 34.
No
interferometric observations of the maser emission have been carried
out so far. These observations would be useful to determine the
excitation source of the maser and to  relate the water maser distribution
with that of the jets in the region.

\subsubsection{CB 54 ([YMT96] CB 54 2)}

CB 54 (LBN 1042), located at 1.5 kpc \citep*{mm-lh97}, 
is an active site of star formation, with
multiple jets \citep*{kha03}
and a bipolar outflow \citep*{out-yc92,out-yc94}. We detected water maser emission towards the
centimeter source [YMT96] CB 54 2 (also named as [YMT96] CB 54 VLA 1; \citealt{cm-y96,cm-m97}).
Within the beam of the
Robledo telescope also fall IRAS 07020-1618, a mm source \citep*{mm-lh97}, a
submillimeter source \citep{smm-l97}, the peak of a CS map \citep*{cs-l98}, and the geometrical center
of the molecular outflow \citep*{out-yc92,out-yc94}. This radio continuum source
is associated with the near-infrared source CB54YC1
\citep*{yun94,yso-yc95}. This source is in its turn composed of at least two
individual near-infrared objects, surrounded by a common nebulosity, and has
been proposed to be an embedded binary system of class I YSOs \citep{yun96}.

The maser emission is highly variable. On 2002 May and 2003 May, the
flux density was below 1 Jy. However, 
a component at $\simeq 7.9$ km s$^{-1}$,
undetected on 2002 May, suffered an outburst in 2003 June-July,
reaching a flux density of $\simeq 50$ Jy. This component was again
undetected (or very weak and blended with a component at 8.7 km s$^{-1}$) on
2005 April.
Such high variability in a water maser is typical of a low-mass
YSO (see e.g., \citealt{wil94}). The velocity of the maser emission is within 15 km s$^{-1}$ from 
the cloud velocity
($V_{\rm LSR}\simeq 19.5$ km s$^{-1}$, CB).

\subsubsection{CB 65 (IRAS 16277-2332)}

CB 65 (LDN 1704) is at a distance of $\sim 160$ pc \citep{smm-v02}, which makes it
the nearest cloud among our
detections. 
We detected maser emission towards IRAS 16277-2332. There is no other
source complying our selection criteria within the Robledo beam around
this IRAS
source. 
Moreover, there is not much known about IRAS
16277-2332, apart from its non-detection in the submillimeter
\citep{smm-v02}, and that 
\citet*{par89} does not include it as associated with
CB 65, so we cannot say much about the nature of
this source. 
Moreover, DG06 did not find high-velocity wings
in the CO spectra towards IRAS 16277-2332 that could indicate the
presence of an outflow from a YSO. This IRAS source was classified in CB as
envelope type (i.e., not within the visible boundaries of the globule) and is
detected only at 60 $\mu$m, with only 0.92 Jy, which is at the detection
limit. 

The maser was detected in 2002 June, but it was not visible in later
observations with the Robledo radio telescope  (Table \ref{tbdetections}), or with the VLA (on 2005 February
12, DG06). The lack of evidence for outflow activity, and the
possible non-association of IRAS 16277-2332 with the globule, suggest
that this source could be an evolved object, although
there are not enough infrared and optical data to support this
suspicion (DG06). However, it is worth noting that the
maser velocity ($V_{\rm LSR} \simeq 1.2$ km s$^{-1}$, Table \ref{tbdetections}) is 
close to the velocity of the cloud
($V_{\rm LSR} \simeq 2.3$ km s$^{-1}$, CB), which would be 
very unlikely if
the object is completely unrelated to the 
globule. 

\subsubsection{CB 101 (IRAS 17503-0833)}

CB 101 (LDN 392) is a globule located at 200 pc
\citep*{lee99}. \citet*{lee99} and \citet*{lee00}
  cataloged it as a starless
core. 

We detected maser emission towards IRAS 17503-0833. No other source with our
selection criteria is found around this IRAS source. The maser does
not show strong variations in flux density, although it showed a
double-peaked profile
in 2002, of which only one component was visible in 2004. The maser
emission, at $V_{\rm LSR}\simeq  29$ km s$^{-1}$ (Table \ref{tbdetections}),
is $\ga 20$ km s$^{-1}$ away from the cloud velocity
($\simeq 6.7$ km s$^{-1}$, CB). 
We note that the centroid velocities of 
water maser emission from YSOs, specially for low-mass sources, is
usually found within $\simeq 15$ km s$^{-1}$ from the cloud 
velocity \citep*{wil94,ang96,
  bra03}. The
high relative velocity of the maser emission in this source with respect to the LSR
velocity of the CB 101 globule, suggests that its pumping source 
may not be associated with the
globule. By inspecting the optical
 images of the Digital Sky Survey, we noted that IRAS 17503-0833 is
out of the optical limit of CB 101, and it should probably not have
been included in the list of IRAS sources of CB.

Due to the lack of high-velocity CO emission and of local enhancement
of molecular gas
towards IRAS 17503-0833, and based on its spectral energy
distribution, DG06 suggested that this source
could be a Mira variable star, rather than a YSO.

\subsubsection{CB 199 ([ARC2001] HH 119 VLA 3)}

CB 199 (B335), located at 250 pc \citep*{tom79}
is an extensively studied site of recent star
formation, which has also been the subject of spectral line studies of
protostellar collapse \citep*{zho93, cho95}.

There is a bipolar molecular outflow in the region \citep*{fre82,gol84}, as well as
the jet-like structure of Herbig-Haro objects HH 199A, B, and C
(\citealt*{vrb86}; \citealt{rei92}).
Both the molecular outflow and the optical jet
are thought to be powered by a far-infrared and submillimeter
source \citep*{kee83,cha90}, which is probably the same object as 
 IRAS 19345+0727.
A total of 13 centimeter sources were detected around this IRAS source
\citep*{cm-a92,cm-a98,cm-a01} of which at least 4 are probably
background objects due to their non-thermal, negative spectral index
at radio wavelenths \citep*{cm-a98}. 

We detected maser emission towards
([ARC2001] HH 119 VLA 3), a  radio continuum source of $\simeq 0.55$ mJy
detected by \citet*{cm-a01}, but undetected in the previous observations
by \citet*{cm-a92} and \citet*{cm-a98}
(upper limit of $\sim 0.15$ mJy), which suggested source variability. 
No other source with our selection criteria
was known within the beam of Robledo from this source. The other
radio continuum sources in the region are more than $2'$ away from this
one. The radio continuum source [ARC92] Barn 335 4 \citep*{cm-a92}, 
associated with IRAS
19345+0727, the powering source of the outflow, lies $\simeq
3\farcm 5$ from [ARC2001] HH 119 VLA 3. Therefore, the water maser seems not
to be related with either this outflow or the HH 199 jet. The lack of
infrared counterpart for [ARC2001] HH 119 VLA 3 in the IRAS point
source and 2MASS catalogs would suggest that this object is
deeply embedded, and thus, very young.

Only one maser component  at $\simeq 37.9$ km s$^{-1}$ is clearly visible in our
spectra. This is $\sim 30$ km s$^{-1}$  from the cloud velocity 
(8.4 km s$^{-1}$, CB), i.e., a shift for the centroid velocity
much larger than expected for masers in YSOs \citep*{wil94,ang96,bra03}. 
Although they are rare, there are some known cases of sources with
large velocity shifts (up to $\simeq 80$ km s$^{-1}$) between water
masers and cloud, specially around HH and GGD objects
\citet*{rod78,rod80}, but they are generally associated with high-mass
YSOs. Large velocity shifts are even rarer
in low-mass objects. If
the maser is associated with a YSO in the globule CB 199, this would
be one of  the cases with the largest velocity offset
between maser emission and cloud velocity know to date for a low-mass
star-forming region. A similar,
unusual shift of $\simeq 30$ km s$^{-1}$ is also found in the low-mass YSO
SVS13 \citep*{cla96}, while values $\simeq 45$ km s$^{-1}$ have been
seen in Cep E-mm \citep*{fur03}, although the latter is probably an
intermediate-mass YSO ($\sim 3$ M$_\odot$;
\citealt{mor01,fro03}). Both SVS13 and Cep E-mm seem to be close
binaries, with separations of $\simeq 65$ AU
\citep*{ang00,eis96}. In the case SVS13, \citet*{rod02} showed that the
high-velocity masers seem to be 
associated with one of the components, although the maser velocities do not seem
to mark the orbital motions in the binary, given the low-mass of the
system. On the other hand, it has been proposed that gravitational
interactions within binaries or multiple systems give rise to
mass-loss outbursts that show up as Herbig Haro jets or FU Ori phenomena
\citep*{rei00,rei04}, and we could speculate that these outbursts  could
also 
show up as high-velocity masers. However, 
it is not possible at this point to determine whether
high-velocity masers are favored in low-mass binary systems, given the scarce
number of objects we are dealing with. Obviously, it would be
interesting to determine whether the source [ARC2001] HH 119 VLA 3 is
indeed a binary.

An alternative possibility to
explain this high-velocity feature is
that [ARC2001] HH 119 VLA 3 is  associated with an
evolved star. 
If this
is true, and the source [ARC2001] HH 119 VLA 3 is the one pumping the
maser, the
presence of radio continuum emission would indicate that it is a
protoplanetary or a young planetary nebula, located behind the CB 199
globule. It is not likely that it is an evolved object 
in the foreground (i.e., closer
than 250 pc), given the lack of an infrared counterpart.

There are no further interferometric observations of the maser in 
this source by
DG65. Given the proximity of CB 199, and the abnormal velocity pattern,
such interferometric observations would be useful to confirm its
association with the radio continuum source and the spatio-kinematical
distribution of the maser emission.

\subsubsection{CB 232 (IRAS 21352+4307)}

CB 232 is a globule located at 600 pc \citep*{mm-lh97}. 
We detected a maser towards IRAS
21352+4307. From the nominal position of this source, 
within the beam of Robledo fall two
submm sources ([HSW99] CB 232 SMM 1 and SMM 2, \citealt{smm-h99}), the
peak of a CS map \citep*{cs-l98}, and the center of a molecular
outflow \citep*{out-yc92,out-yc94}. 

The maser emission shows a varying velocity pattern. The maximum
emission was at
$\simeq 12.1$ km s$^{-1}$ in 2003 May, although a weaker component at
$\simeq 10$  km s$^{-1}$ was also present. This weaker component was
the dominant one in later spectra, while the one at 12.1  km s$^{-1}$
was absent. The detected maser components are 
close to the cloud velocity ($\simeq 12.6$   km s$^{-1}$, CB).

The variation of the maser spectrum
and the presence of other signs of star
formation indicates that the maser emission is pumped by a low-mass
YSO. 
There is a near-infrared infrared source (CB232YC1-I),
classified as a class I object by \citet*{yso-yc95}, and located between
SMM 1 and SMM 2, and at $\simeq$ 10''
and $\simeq$ 5'' from each of them, respectively. The identification
of CB232YC1-I with SMM 2 is uncertain, 
given the positional
uncertainty of the latter ($\simeq4''$ \citealt{smm-h99}). Therefore,
there are at least 2 objects within $10''$. 
Further interferometric observations of the water
maser (DG06) found this emission to lie at the position of CB232YC1-I, which
confirms that this class I source could be the powering source of the
molecular outflow. 

\section{Discussion}
\label{sec:discussion}

Of the selection criteria used to select our target sources, the
highest detection rates of masers are those related to the presence of molecular
outflows or the peak in maps of the high-density tracer CS. We
performed a statistical study to try to determine whether any of the
selection criteria is a better predictor for the presence of water
maser emission. For this, we used a Fisher's Exact test and a 95\%
confidence level. This test indicates that the presence of a
submillimeter source, of a peak
of CS, or of a molecular outflow are all a better predictor of the
existence of water maser emission than the 
presence of an IRAS source, which is reasonable, since the latter was
the less restrictive of all our criteria in the identification of YSOs
candidates. Apart for IRAS sources, the detection rates for all other selection criteria (even
the 0\% rate for peaks of NH$_3$ maps) are all statistically compatible.

Another relevant fact is that most of the globules in the CB catalog
which are known to be associated with water masers 
(CB 3, CB 34, CB 39, CB 54, CB 205) are located at distances $\ge 1$
kpc.  Therefore, they tend to be larger than probably intended by CB
in their catalog, which tried to select ``small'' (size $\la 1.5$ pc)
globules, using a criterion of $< 10'$ assuming distances 
$\la 500$ pc. So the globules mentioned above belong to a class of
``large'' Bok globules. Large, massive globules are
obviously more likely to be sites of star formation, with a larger number of
YSOs, and therefore, with a higher probability of hosting at least one
water-maser-emitting object. YSOs in larger globules would also tend to be
more massive  (as in the case of CB 3) than in the case
of smaller globules, which also increases the probability of the presence of
water masers. 
In our
survey we have identified the smallest Bok globules
known to harbor water maser emission. In particular, CB 232, whose maser
is most likely associated with a YSO, has a size of $\simeq 0.6$
pc ($5\farcm 6\times 2\farcm 2$ at 600 pc; CB; \citealt{mm-lh97}).
The cases of CB 65 ($\simeq 0.3$ pc size) and CB 199 ($\simeq 0.5$ pc size) 
need further investigation to determine the nature of their
pumping sources, but these three sources seem to be the best candidates
for further use of water masers as a tool to study star formation in
small Bok globules.

\section{Conclusions}
\label{sec:conclusions}
In this paper, we present the most sensitive survey for water maser emission
towards Bok globules to date, using NASA's 70m antenna
in Robledo de Chavela (Spain). A total of 203 target positions within
the clouds of the \citet{cb} catalogs were observed. Our main results are as follow:
\begin{itemize}
\item We have obtained 7 maser detections, 6 of which (in CB 34, CB
  54, CB 65, CB 101, CB 199, and CB 232) are new. Of the previously
  known masers in the CB catalog, only the one in CB 3 was detected by
  us. No emission was seen towards the previously reported masers in 
either CB 39 or CB 205.
\item Of our detections, the ones in CB 3, CB 34, CB 54, and CB 232
  are most likely associated with YSOs. We suggest that the maser in
  CB 101 is associated with an evolved object. The nature of CB 65 is
  uncertain.
\item In the case of CB 199, the relatively large shift ($\simeq 30$
  km s$^{-1}$) of
  the centroid velocity of the maser emission with respect to the
  cloud velocity is unusual for YSOs, specially for low-mass ones. We
  speculate that the presence of high-velocity masers in low-mass
  YSOs might be related with episodes of energetic mass loss in close
  binaries. Alternatively, the maser in CB 199 could be related to a
  protoplanetary or a young planetary nebula.
\item CB 232 is the smallest Bok globule (size $\simeq 0.6$ pc) 
  known to be associated with
  water maser emission. However, if their association with YSOs is
  confirmed, CB 65 (size $\simeq 0.3$ pc) and CB 199 ($\simeq 0.5$ pc)
  are even smaller globules. These objects are good candidates for the
  study of relatively isolated star formation with high spatial
  resolution.
\item Of our selection criteria, the more restrictive ones for the
  identification of YSOs (radio continuum emission, peak of a
  high-density tracer, and geometrical center of a molecular outflow)
  show statistically compatible rates of water maser detection. Only
  the presence of IRAS sources tends to be a somewhat worse predictor for the
  presence of masers.
\end{itemize}

\acknowledgments

We are deeply indebted to many staff members at the Madrid Deep Space
Communication Complex and Jet Propulsion Laboratory, without
whose invaluable help spectroscopy observations at Robledo (and
therefore this paper) would have never been possible. We would also
like to
thank the water megamaser group at the Harvard-Smithsonian Center for Astrophysics  (Lincoln Greenhill, Paul Kondratko, and James Moran)
for allowing us to use their SAO-4K spectrometer during four nights
in which our 256-channel spectrometer was not working. We also thank
Paul Kondratko 
for providing us with his SAO-4K data processing software, which was useful
to develop ours for the 384-channel spectrometer (SPB500).
JFG acknowledges support from MEC (Spain) grant
AYA 2005-08523-C03-03 and from Junta de Andaluc\'{\i}a (TIC-126). OS is
partially supported by MEC grant AYA2003-09499. IdG acknowledges the
support of a Calvo Rod\'es predoctoral fellowship from the Instituto
Nacional de T\'ecnica Aeroespacial during the
development of this work.
This paper is based on observations taken during
``host-country'' allocated time at Robledo de Chavela; this time is managed
by the LAEFF of INTA, under agreement with NASA/INSA. It also makes
use of data products from the Two Micron All Sky Survey (2MASS), which
is a joint project of the University of Massachusetts and the Infrared
Processing and Analysis Center at the California Institute of
Technology, funded by NASA and the NSF. We acknowledge the use 
of the free GILDAS software of IRAM (http://www.iram.fr/IRAMFR/GILDAS) 
for data reduction of our spectral line data.  This research has also made use
of 
the SIMBAD database, 
operated at CDS, Strasbourg, France.

\clearpage

\clearpage

\clearpage

\clearpage

\clearpage

\clearpage

\clearpage

\clearpage

\begin{deluxetable}{llllll}
\tabletypesize{\scriptsize}
\rotate
\tablecaption{Sources searched for water maser emission\label{tbobserved}}
\tablewidth{0pt}
\tablehead{
\colhead{Globule} & \colhead{Source\tablenotemark{a}} & \colhead{Right Ascension\tablenotemark{b}}   &
\colhead{Declination\tablenotemark{b}}   & 
\colhead{Included Sources\tablenotemark{c}} &
\colhead{References\tablenotemark{d}} \\
\colhead{} & \colhead{} & \colhead{(J2000)} & \colhead{(J2000)} & \colhead{} & 
\colhead{}
}

\startdata
CB 3  & [YMT96] CB 3 1  & 00 28 22.0 & +56 41 39 & & 6 
                                                         \\
      & CB3-mm\tablenotemark{e}& 00 28 42.7 & +56 42 06 & smm ([HSW99] CB 3
SMM 1), IRAS 00259+5625, outf, cs & 1, 
                   2, 
                   3, 
                   4, 
                   5 
                   \\
CB 4  & IRAS 00362+5234 & 00 39 03.5 & +52 50 57 & \\
CB 6  & IRAS 00465+6028 & 00 49 25.0 & +50 44 45 & mm, smm
                                                             & 1,
                                                               2 \\
CB 7  & IRAS 01078+6409 & 01 11 03.4 & +64 25 24 & \\
      & IRAS 01087+6404 & 01 11 58.0 & +64 20 07 & \\
CB 8  & IRAS 01202+7406 & 01 24 12.4 & +74 22 03 & \\
CB 11 & IRAS 01333+6448 & 01 36 51.5 & +65 03 55 & \\
      & IRAS 01334+6442 & 01 36 59.4 & +64 57 39 & \\
      & IRAS 01341+6447 & 01 37 41.3 & +65 03 08 & \\
CB 12 & CS peak         & 01 38 32.7 & +65 06 02 &  & 5\\
      & IRAS 01354+6447 & 01 38 56.8 & +65 03 12 & \\
CB 15 & IRAS 03521+5555 & 03 56 07.9 & +56 04 30 & \\
      & IRAS 03523+5608 & 03 56 21.7 & +56 17 01 & 
        \\
      & IRAS 03535+5555 & 03 57 30.4 & +56 04 15 & 
         \\
CB 16 & IRAS 03592-5642 & 04 03 15.6 & +56 50 27 & \\
CB 17 & IRAS 04005+5647 & 04 04 33.7 & +56 56 10 & mm, smm & 1, 2\\
      & NH$_3$ peak     & 04 04 38.0 & +56 56 11 & & 7 
   \\ 
CB 19 & IRAS 04233+2529 & 04 26 21.3 & +25 36 22 & \\
      & IRAS 04240+2535 & 04 27 02.7 & +25 42 24 & 
                                                   \\
CB 22 &  NH$_3$ peak    & 04 40 33.2 & +29 55 04 & & 7\\ 
CB 23 & NH$_3$ peak     & 04 43 30.0 & +29 39 01 & & 7\\  
      & CB23-cm1       & 04 43 34.9 & +29 38 05 & & 8\\
      & CB23-cm2       & 04 43 35.0 & +29 37 13 & & 8\\
CB 26 & IRAS 04559+5200 & 04 59 52.4 & +52 04 45 & mm 
                                                & 1\\
CB 27 & IRAS 05013+3234 & 05 04 37.2 & +32 38 15 & \\
CB 28 & IRAS 05036-0359 & 05 06 08.9 & $-03$ 55 16 & \\
      & IRAS 05037-0402 & 05 06 13.7 & $-03$ 58 60 & \\
      & CS peak         & 05 06 19.4 & $-03$ 56 24 & IRAS 05038-0400 \\
      & IRAS 05038-0400 & 05 06 19.9 & $-03$ 56 33 & cs & 5\\
      & NH$_3$ peak     & 05 06 20.2 & $-03$ 56 02 & & 7\\
CB 29 & IRAS 05190-0348 & 05 21 31.5 & $-03$ 45 08 & \\
      & IRAS 05194-0346 & 05 21 56.2 & $-03$ 44 08 & & 5\\
      & IRAS 05194-0343 & 05 21 57.5 & $-03$ 40 35 & \\
      & CS peak         & 05 22 10.3 & $-03$ 41 06 & & 5\\
      & IRAS 05201-0341 & 05 22 38.5 & $-03$ 38 53 & \\
CB 30 & [MYT99] CB 30 3 & 05 29 30.5 & +05 43 22 & &9\\
      & IRAS 05268+0550 & 05 29 31.7 & +05 52 55 & \\
      & IRAS 05268+0538 & 05 29 32.2 & +05 40 34 & cm ([MYT99] CB30
4), cs & 9, 5\\
      & IRAS 05274+0542 & 05 30 06.7 & +05 44 23 & \\
CB 31 & [YMT96] CB 31 1 & 05 33 14.5 & $-00$ 35 34 & & 6\\
      & [YMT96] CB 31 2 & 05 33 15.8 & $-00$ 35 20 & & 6\\
      & IRAS 05307-0038 & 05 33 18.2 & $-00$ 36 13 & \\
CB 32 & IRAS 05344-0016 & 05 36 59.7 & $-00$ 14 21 & \\
      & CS peak         & 05 38 28.2 & $-00$ 17 25 & & 5\\
CB 33 & IRAS 05433+2040 & 05 46 17.7 & +20 41 39 & \\
      & IRAS 05437+2045 & 05 46 41.1 & +20 46 10 & \\
CB 34 & [HSW99] CB 34 SMM 2 & 05 47 00.0 & +21 00 30 & & 3\\
      & [YMT96] CB 34 3 & 05 47 01.7 & +21 00 25 & mm, smm ([HSW99] CB
34 SMM 1), IRAS 05440+2059, outf, cs, nh3 & 6, 1, 2, 3, 4, 5, 10\\
      & [HSW99] CB 34 SMM 1 & 05 47 01.9 & +21 00 06 & cm ([YMT96] CB
      34 3), IRAS 05440+2059 & 3, 6\\
      & [HSW99] CB 34 SMM 3 & 05 47 05.3 & +21 00 42 & smm ([HSW99] CB
      34 SMM 4) & 3\\
      & [HSW99] CB 34 SMM 5 & 05 47 07.8 & +21 00 04 &  & 3\\
CB 37 & [MYT99] CB 37 1 & 06 00 35.6 & +31 37 37 & & 11\\
      & NH$_3$ peak     & 06 00 37.3 & +31 38 50 & & 7\\
      & [MYT99] CB 37 2 & 06 00 47.6 & +31 40 48 & & 11\\
CB 39 & IRAS 05591-1630\tablenotemark{f} & 06 01 59.5 & +16 30 56 &
outf & 21
                                    \\
      & CS peak         & 06 01 59.6 & +16 31 44 & & 5\\
      & [YMT96] CB 39 2 & 06 02 04.6 & +16 30 06 & & 6\\
CB 40 & IRAS 05589+1638 & 06 01 53.1 & +16 38 35 & \\
      & IRAS 05592+1640 & 06 02 07.2 & +16 40 11 & \\
CB 42 & IRAS 06000+1644 & 06 02 55.7 & +16 44 36 & \\
CB 43 & [MYT99] CB 43 1 & 06 03 10.6 & +16 36 07 & & 11\\
CB 44 & CS peak         & 06 07 29.3 & +19 27 53 & & 5\\
      & [MYT99] CB 44 2 & 06 07 37.2 & +19 25 12 & & 11\\
      & [MYT99] CB 44 3 & 06 07 37.6 & +19 28 35 & & 11\\
      & IRAS 06047+1923 & 06 07 39.2 & +19 22 43 & \\
      & IRAS 06048+1934 & 06 07 47.4 & +19 34 18 & \\
CB 45 & IRAS 06055+1800 & 06 08 27.1 & +17 59 59 & \\
      & NH$_3$ peak     & 06 08 56.6 & +17 50 19 & & 7\\
CB 48 & IRAS 06175+0711 & 06 20 15.6 & +07 10 14 & \\
CB 50 & IRAS 06316+0748 & 06 34 19.1 & +07 45 46 & \\
CB 51 & IRAS 06355+0134 & 06 38 07.3 & +01 31 56 & \\
CB 52 & [YMT96] CB 52 1 & 06 48 33.5 & $-16$ 50 31 & & 6\\
      & IRAS 06464-1650 & 06 48 39.2 & $-16$ 54 04 & mm, smm & 1, 2\\
      & IRAS 06464-1644 & 06 48 41.8 & $-16$ 48 06 & \\
      & IRAS 06471-1651 & 06 49 22.2 & $-16$ 55 01 & \\
CB 54 & [YMT96] CB 54 2 & 07 04 21.2 & $-16$ 23 15 & mm, smm, IRAS
07020-1618, outf, cs & 6, 1, 2, 4, 5\\
CB 55 & IRAS 07019-1631 & 07 04 12.6 & $-16$ 35 34 & \\
CB 56 & IRAS 07125-2507 & 07 14 36.5 & $-25$ 12 57 & \\
      & IRAS 07125-2503 & 07 14 38.9 & $-25$ 08 54 & \\
CB 57 & IRAS 07154-2304 & 07 17 32.2 & $-23$ 09 42 &\\
      & IRAS 07156-2248 & 07 17 45.5 & $-22$ 54 14 &\\
      & IRAS 07156-2301 & 07 17 47.8 & $-23$ 07 01 &\\
CB 58 & IRAS 07159-2329 & 07 18 02.5 & $-23$ 35 06 & \\
      & [YMT96] CB 58 1 & 07 18 07.6 & $-23$ 41 07 & & 6\\
      & IRAS 07161-2336 & 07 18 15.2 & $-23$ 41 42 & mm, smm & 1, 2\\
CB 59 & IRAS 07171+0359 & 07 19 44.6 & +03 53 48 & \\
CB 60 & IRAS 08022-3115 & 08 04 11.9 & $-31$ 24 00 & \\
      & IRAS 08026-3122 & 08 04 36.9 & $-31$ 30 44 & \\
      & IRAS 08029-3118 & 08 04 56.4 & $-31$ 27 23 & \\
CB 63 & IRAS 15486-0350 & 15 51 16.7 & $-03$ 59 41 & \\
CB 65 & IRAS 16277-2332 & 16 30 43.7 & $-23$ 39 08 & \\
      & [VRC2001] L1704 SMM 1  & 16 30 50.6 & $-23$ 42 08 & & 12\\
      & IRAS 16287-2337 & 16 31 44.6 & $-23$ 43 55 & \\
CB 67 & IRAS 16485-1906 & 16 51 10.3 & $-19$ 11 12 & \\
CB 68 & CS peak         & 16 57 16.3 & $-16$ 07 40 & & 5\\
      & IRAS 16544-1604\tablenotemark{g} & 16 57 19.5 & $-16$ 09 25 &
mm, smm ([HSW99] CB 68 SMM 1), outf & 1, 3, 13\\
      & NH$_3$ peak     & 16 57 20.5 & $-16$ 09 02 & & 7\\
CB 78 & IRAS 17147-1821 & 17 17 38.5 & $-18$ 24 14 & \\
CB 81 & outflow         & 17 22 26.5 & $-27$ 08 10 & & 4\\
CB 82 & [VRC2001] L57 SMM 1       & 17 22 38.5 & $-23$ 49 57 & & 12\\
      & NH$_3$ peak     & 17 22 39.3 & $-23$ 49 46 & & 14\\
      & CS peak         & 17 22 40.7 & $-23$ 48 46 & & 5\\
CB 98 & CB98-mm         & 17 47 00.7 & $-20$ 30 29 & & 1\\
CB 100& IRAS 17490-0258 & 17 51 38.1 & $-02$ 59 07 &\\
      & IRAS 17493-0258 & 17 52 00.4 & $-02$ 58 59 &\\
CB 101& IRAS 17503-0833 & 17 53 05.2 & $-08$ 33 41 & \\
      & IRAS 17505-0828 & 17 53 14.1 & $-08$ 28 42 &\\
CB 104& IRAS 17526-0815 & 17 55 21.1 & $-08$ 15 38 &\\
      & IRAS 17533-0808 & 17 56 03.7 & $-08$ 09 16 &\\
CB 105& IRAS 17561-0349 & 17 58 44.5 & $-03$ 49 52 &\\
CB 106& IRAS 17577-0329 & 18 00 25.8 & $-03$ 29 35 &\\
      & IRAS 17580-0330 & 18 00 41.4 & $-03$ 30 30 &\\
      & IRAS 17586-0330 & 18 01 18.3 & $-03$ 30 01 &\\
CB 108& [YMT96] CB 108 1& 18 03 01.9 & $-20$ 49 37 & & 6\\
CB 118& IRAS 18094-1550 & 18 12 22.1 & $-15$ 49 13 &\\
CB 121& IRAS 18115-0701 & 18 14 17.6 & $-07$ 00 51 &\\
      & IRAS 18116-0657 & 18 14 18.9 & $-06$ 56 44 & \\
CB 124& IRAS 18120+0704 & 18 14 31.4 & +07 05 13 &\\
      & IRAS 18122+0703 & 18 14 42.7 & +07 04 44 &\\
CB 125& NH$_3$ peak     & 18 15 34.7 & $-18$ 11 12 & & 7\\
CB 128& IRAS 18132-0350 & 18 15 52.5 & $-03$ 49 40 & \\
CB 130& CB130-mm        & 18 16 14.8 & $-02$ 32 47 & nh3 & 1, 7\\
CB 131& CB131-smm       & 18 17 00.5 & $-18$ 02 04 & & 2\\
CB 137& IRAS 18219-0100 & 18 24 30.3 & $-00$ 58 22 &\\
CB 142& IRAS 18272-1343 & 18 30 02.0 & $-13$ 41 07 &\\
CB 145& IRAS 18296-0911 & 18 32 21.7 & $-09$ 09 25 & mm & 1\\
CB 146& IRAS 18293-0906 & 18 32 08.2 & $-09$ 03 57 & \\
      & IRAS 18294-0901 & 18 32 11.1 & $-08$ 59 05 & \\
      & [HSW99] CB 146 SMM 2& 18 32 19.4 & $-08$ 53 00 & & 15\\
      & [HSW99] CB 146 SMM 1& 18 32 21.3 & $-08$ 51 56 & & 15\\
CB 171& [YMT96] CB 171 2& 19 01 33.6 & $-04$ 31 48 & & 6\\
      & [YMT96] CB 171 3& 19 01 55.8 & $-04$ 31 11 & & 6\\
CB 177& IRAS 18599+1739 & 19 02 07.7 & +17 43 58 & \\
CB 178& IRAS 18595+1812 & 19 01 44.0 & +18 16 29 & \\
      & IRAS 19002+1755 & 19 02 28.0 & +17 59 58 & \\
CB 180& NH$_3$ peak     & 19 06 08.6 & $-06$ 52 47 & & 7\\
CB 184& IRAS 19116+1623 & 19 13 56.4 & +16 28 27 & \\
CB 187& IRAS 19162-0135 & 19 18 48.7 & $-01$ 29 39 & \\
CB 188& IRAS 19179+1129 & 19 20 14.9 & +11 35 35 & mm & 1\\
      & Outflow         & 19 20 16.3 & +11 35 57 & & 4\\
      & CS peak         & 19 20 19.9 & +11 35 57 & & 5\\
      & IRAS 19180+1127 & 19 20 21.0 & +11 32 54 & \\
CB 189& IRAS 19180+1116 & 19 20 22.5 & +11 22 07 & \\
      & [VRC2001] L673 SMM 7& 19 20 23.1 & +11 22 50 & & 12\\
      & [VRC2001] L673 SMM 1& 19 20 25.2 & +11 22 17 & & 12\\
      & IRAS 19180+1114 & 19 20 25.8 & +11 19 52 & smm ([VRC2001] L673
      SMM 2), outf & 12, 15  \\
      & IRAS 19183+1123 & 19 20 44.0 & +11 28 55 & \\
      & IRAS 19184+1118 & 19 20 45.6 & +11 23 50 & \\
CB 190& IRAS 19186+2325 & 19 20 46.3 & +23 31 31 & \\
CB 194& IRAS 19273+1433 & 19 29 35.7 & +14 39 23 & \\
CB 196& IRAS 19329+1213 & 19 35 18.0 & +12 20 36 & \\
CB 198& IRAS 19342+1213 & 19 36 37.8 & +12 19 59 & \\
CB 199& [ARC92] Barn 335 9& 19 36 44.4 & +07 36 43 & & 17\\
      & [ARC92] Barn 335 1& 19 36 47.6 & +07 32 58 & & 17\\
      & IRAS 19343+0727 & 19 36 48.8 & +07 34 29 & \\
      & [ARC92] Barn 335 2& 19 36 49.5 & +07 35 04 & & 17\\
      & NH$_3$ peak     & 19 36 59.0 & +07 34 17 & & 19\\
      & IRAS 19345+0727 & 19 37 01.0 & +07 34 11 & cm ([ARC92] Barn
      335 4), smm ([HSW99] B 335 SMM 1), outf & 17, 2, 18\\ 
      & [ARC92] Barn 335 11& 19 37 08.8 & +07 31 41 & & 17\\
      & [ARC2001] HH 119 VLA 3 & 19 37 10.2 & +07 36 50 & & 20\\
      & IRAS 19347+0729 & 19 37 10.5 & +07 36 26 & \\
      & IRAS 19348+0724 & 19 37 17.0 & +07 31 57 & \\
CB 203& IRAS 19413+1902 & 19 43 33.5 & +19 09 52 & \\
CB 205& IRAS 19427+2741 & 19 44 45.5 & +27 48 37 & \\ 
      & [YMT96] CB 205 1& 19 45 09.5 & +27 51 06 & & 6\\
      & [HSW99] L 810 SMM 2& 19 45 21.3 & +27 50 40 & & 2\\ 
      & [YMT96] CB 205 2& 19 45 21.9 & +27 53 40 & & 6\\
      & IRAS 19433+2743\tablenotemark{h} & 19 45 23.9 & +27 50 58 &
mm, smm ([HSW99] L 810 SMM 1), outf, nh3 & 1, 2, 21, 22\\
      & IRAS 19433+2751 & 19 45 25.2 & +27 58 54 & \\
      & [YMT96] CB 205 3& 19 45 35.1 & +27 54 11 & & 6\\   
      & IRAS 19438+2757 & 19 45 51.2 & +28 04 23 & \\
      & IRAS 19438+2737 & 19 45 55.5 & +27 44 58 & \\
      & IRAS 19439+2748 & 19 45 57.8 & +27 56 06 & \\
CB 206& [YMT96] CB 206 2& 19 46 30.4 & +19 06 06 & & 6\\
CB 207& IRAS 19437+2108 & 19 45 53.0 & +21 16 15 & \\
CB 208& IRAS 19450+1847 & 19 47 14.7 & +18 55 10 & \\
CB 210& IRAS 19529+3341 & 19 54 49.1 & +33 49 15 & \\
CB 211& IRAS 19576+2447 & 19 59 46.5 & +24 55 35 & \\
CB 214& IRAS 20018+2629 & 20 03 57.9 & +26 38 20 & outf & 4, 21\\
CB 216& Outflow         & 20 05 49.7 & +23 27 04 & &4\\
      & IRAS 20037+2317 & 20 05 53.6 & +23 26 34 & cs & 5\\ 
      & CS peak         & 20 05 54.1 & +23 26 16 & IRAS 20037+2317 & 5\\
CB 217& Outflow         & 20 07 45.9 & +37 07 01 & & 21\\
      & [YMT96] CB 217 4& 20 07 51.5 & +37 06 55 & & 6\\            
CB 219& IRAS 20176+6343 & 20 18 23.7 & +63 52 30 & \\
CB 222& IRAS 20328+6351 & 20 33 36.4 & +64 02 21 & \\
CB 224& NH$_3$ peak     & 20 36 20.3 & +63 52 55 & & 7 \\
      & IRAS 20355+6343 & 20 36 22.1 & +63 53 39 & mm & 1\\
CB 225& IRAS 20365+5607 & 20 37 48.0 & +56 17 56 & \\
CB 230& [YMT96] CB 230 1& 21 17 30.9 & +68 18 10 & \\
      & [YMT96] CB 230 2& 21 17 38.5 & +68 17 32 & mm, smm ([HSW99] CB
      230 SMM 1), IRAS 21169+6804, outf, cs & 6, 1, 2, 15, 4, 21, 5\\
CB 232& IRAS 21352+4307 & 21 37 11.3 & +43 20 36 & smm ([HSW99] CB 232
SMM 1, [HSW99] CB 232 SMM 2), outf, cs & 15, 4, 5\\
CB 233& CS peak         & 21 40 26.4 & +57 48 06 & & 5\\
CB 235& IRAS 21548+5843 & 21 56 25.7 & +58 57 54 & \\
CB 238& NH$_3$ peak & 22 13 22.6 & +41 02 51 & & 7\\
CB 240& IRAS 22317+5816 & 22 33 39.3 & +58 31 56 & mm & 1 \\
CB 241& IRAS 23095+6547 & 23 11 37.0 & +66 04 07 & \\
CB 243& IRAS 23228+6320 & 23 25 05.7 & +63 36 34 & mm, smm ([VRC2001]
L1246 SMM 1), nh3 & 1, 23, 12, 7\\
      & NH$_3$ peak     & 23 25 06.9 & +63 36 50 & smm ([VRC2001]
      L1246 SMM 1), IRAS 23228+6320 & 7, 23, 12\\
      & [VRC2001] L1246 SMM 2& 23 25 16.4 & +63 36 46 & & 23, 12\\
      & CS peak         & 23 25 27.0 & +63 35 46 & & 5\\
CB 244 &[VRC2001] L1262 SMM 2& 23 25 26.0 & +74 18 28 & nh3 & 12, 23, 24\\
       & [YMT96] CB 244 1& 23 25 46.6 & +74 17 40 & mm, smm ([VRC2002]
L1262 SMM 1), IRAS 23238+7401, outf & 6, 1, 2, 12, 4 \\
       & IRAS 23249+7406 & 23 26 53.2 & +74 22 34 & \\
CB 246& CB246-mm        & 23 56 43.6 & +58 34 29 & & 1\\
CB 247& IRAS 23550+6430 & 23 57 36.4 & +64 46 48 & \\
\enddata

\tablenotetext{a}{Target sources. SIMBAD names were used, where
  available. Labels [YMT96], [MYT99], [ARC92], and
  [ARC2001] indicate cm sources}
\tablenotetext{b}{Coordinates of pointing position. Units of right ascension are
  hours, minutes, and seconds. Units of declination are degrees,
 arcminutes, and arcseconds}
\tablenotetext{c}{Other sources included within the telescope beam, complying
  any of the selection criteria mentioned in
  sec. \ref{sec:sample}. Cm: centimeter source; mm: millimeter source;
  smm: submillimeter source; outf: center of molecular outflow; nh3:
  peak of NH$_3$ map; cs: peak of CS map. Where available, SIMBAD names are given
  between parentheses}
\tablenotetext{d}{References for the sources complying the selection
  criteria (except for IRAS sources)}
\tablenotetext{e}{Maser detected by \citet{sca91}}
\tablenotetext{f}{Maser detected by \citet*{sch75}}
\tablenotetext{g}{Coordinates used for IRAS 16544-1604 are those in
  the IRAS Point Source Catalog. SIMBAD reports for this source the
  coordinates of F16544-1604 in the IRAS Faint Source Catalog, which
  is also within the Robledo beam from our pointing position}
\tablenotetext{h}{Maser detected by \citet*{nec85}}
\tablerefs{(1) \citet*{mm-lh97};
                   (2)  \citet{smm-l97};
                   (3) \citet{smm-h00};
                   (4) \citet*{out-yc92}; 
                   (5) \citet*{cs-l98};
                   (6) \citet*{cm-y96};
                   (7) \citet*{lem96};
                   (8) \citet*{har02}; 
                   (9) \citet*{cm-m99}; 
                   (10) \citet*{cod98};
                   (11) \citet*{cm-m99};
                   (12) \citet{smm-v02};
                   (13) \citet*{val00};
                   (14) \citet*{bou95};
                   (15) \citet{smm-h99};
                   (16) \citet*{arm89};
                   (17) \citet*{cm-a92};
                   (18) \citet*{fre82};
                   (19) \citet*{ben89};
                   (20) \citet*{cm-a01};
                   (21) \citet*{out-yc94};
                   (22) \citet*{nec85};
                   (23) \citet*{smm-v01}; 
                   (24) \citet*{ben84}
                  }
\end{deluxetable}

\begin{deluxetable}{lllllrl}
\tabletypesize{\scriptsize}
\tablecaption{Water maser detections\label{tbdetections}}
\tablewidth{0pt}
\tablehead{
\colhead{Globule} & \colhead{Source} & 
\colhead{$S_\nu$\tablenotemark{a}} & \colhead{$\int S_\nu\; dV$\tablenotemark{b}} &
\colhead{$V_{\rm peak}$\tablenotemark{c}} &
\colhead{$V_{\rm cloud}$\tablenotemark{d}} &
\colhead{Date\tablenotemark{e}} \\
\colhead{} & \colhead{} &  
\colhead{(Jy)} & \colhead{(Jy km s$^{-1}$)} &
\colhead{(km s$^{-1}$)} &
\colhead{(km s$^{-1}$)} &
\colhead{} 
}

\startdata
CB 3  & CB3-mm          & $11.0\pm 0.4$ & $52\pm 3$ & $-53.4\pm 0.6$ &
$-38.3$ & 2004-JUL-24\\ 
      &                 & $12.4\pm 0.4$ & $80.2\pm 2.4$ & $-53.5\pm 0.6$ &
& 2004-AUG-19\\ 
      &                 & $19.60\pm 0.22$ & $97\pm 3$ & $-41.1\pm 0.6$ &
& 2005-JAN-03\\ 
      &                 & $12.00\pm 0.13$ & $56.0\pm 0.7$ & $-41.8\pm 0.6$ &
& 2005-JAN-31\\ 
CB 34 & [HSW99] CB 34 SMM 3          & $0.34\pm 0.08$ & $0.57\pm 0.23$
& $8.1\pm 0.6$  &
0.7 & 2004-AUG-20\\ 
      &                 & $0.33\pm 0.07$ & $0.57\pm 0.20$ & $8.1\pm 0.6$ &
& 2004-AUG-25\\ 
      &                 & $<0.14$ &                         &     &
& 2005-APR-14\\ 
      &                 & $<0.14$ &                         &     &
& 2005-APR-18\\ 
CB 54 & [YMT96] CB 54 2   & $0.84\pm 0.22$ & $1.1\pm 0.5$ 
& $13.7\pm 0.5$ & 
19.5 & 2002-MAY-26\\ 
      &                 & $0.88\pm 0.20$ & $0.7\pm 0.3$ 
& $7.9\pm 0.5$ & 
& 2003-MAY-04\\ 
      &                 & $14.1\pm 0.4$ & $12.6\pm 0.7$ 
& $7.9\pm 0.5$ & 
& 2003-JUN-07\\ 
      &                 & $49.3\pm 0.9$ & $45.3\pm 1.8$ 
& $7.9\pm 0.5$ & 
& 2003-JUN-29\\ 
      &                 & $41.1\pm 0.5$ & $39.9\pm 0.9$ 
& $7.9\pm 0.5$ & 
& 2003-JUL-01\\ 
      &                 & $5.24\pm 0.11$ & $9.5\pm 0.4$ & $8.7\pm 0.6$ &
& 2005-APR-02\\ 
CB 65 & IRAS 16277-2332 & $0.30\pm 0.19$ & $0.6\pm 0.3$ 
& $1.2\pm 0.5$ & 
2.3 & 2002-JUN-16\\ 
      &                 & $<0.25$  &                         &     &
& 2004-JUL-08\\ 
      &                 & $<0.18$  &                         &     &
& 2004-JUL-14\\ 
      &                 & $<0.22$  &                         &     &
& 2005-JUN-05\\ 
CB 101 & IRAS 17503-0833 & $0.26\pm 0.05 $ &  $0.41\pm 0.13$  
& $29.3\pm1.3$ & 6.7 & 2002-APR-10\\
       &                 & $0.32\pm 0.09$ & $0.41\pm 0.15$ 
& $28.3\pm 0.5$ & 
& 2002-APR-14\\ 
      &                  & $0.55\pm 0.19$ & $0.63\pm 0.25$ 
& $29.3\pm 0.5$ &
& 2002-NOV-15\\ 
       &                & $0.60\pm 0.15$ & $0.52\pm 0.21$ & $28.9 \pm 0.6$&
& 2004-JUL-07\\ 
       &                & $0.34\pm 0.17$ & $0.6\pm 0.4$  & $28.9\pm 0.6$ &
& 2004-JUL-14\\ 
       &                & $0.58\pm 0.18$ & $0.6\pm 0.3$  & $29.0\pm 0.6$ &
& 2004-AUG-04\\ 
CB 199 & [ARC2001] HH 119 VLA 3 & $0.49\pm 0.06$ & $0.69\pm 0.15$ &
$37.9\pm 0.6$ & 8.4 & 2005-JUN-16 \\
       &                        & $0.36\pm 0.10$ & $0.41\pm 0.18$ & 
$37.8\pm 0.6$ & & 2005-JUL-21 \\
CB 232 &   IRAS 21352+4307   & $0.34\pm 0.10$ & $0.46\pm 0.19$ &
$12.1\pm 0.5$ &
12.6 & 2003-MAY-10\\ 
       &                & $0.82\pm 0.13$ & $1.10\pm 0.23$ & $12.1\pm 0.5$ &
& 2003-MAY-11\\ 
       &                & $0.62\pm 0.19$ & $0.51\pm 0.24$ & $10.5\pm 0.5$ &
& 2003-JUL-15\\ 
       &                & $0.94\pm 0.12$ & $0.95\pm 0.23$ & $9.4\pm 0.6$ &
& 2005-APR-02\\ 

\enddata
\tablenotetext{a}{Flux density of the strongest maser
  feature. Uncertainties are $2\sigma$.}
\tablenotetext{b}{Flux density integrated over the velocity extent of
  the maser emission. Uncertainties are $2\sigma$.}
\tablenotetext{c}{LSR Velocity of the strongest maser feature}
\tablenotetext{d}{LSR Velocity of the globule, as given in the CB catalog}
\tablenotetext{e}{Date of observation}

\end{deluxetable}

\begin{deluxetable}{llllll}
\tablecaption{Non detections \label{tbnondetections}}
\tablewidth{0pt}
\tablehead{
\colhead{Globule} & \colhead{Source} & 
\colhead{$V_{\rm min}$\tablenotemark{a}} & 
\colhead{$V_{\rm max}$\tablenotemark{a}} & 
\colhead{Rms\tablenotemark{b}}   &
\colhead{Date\tablenotemark{c}}  \\
\colhead{} & \colhead{} & 
\colhead{(km s$^{-1}$)} & 
\colhead{(km s$^{-1}$)} &
\colhead{(Jy)} & \colhead{}
}

\startdata
CB 3  & [YMT96] CB 3 1  & -2696 & 2698 & 0.03 & 2002-APR-09\\
      &                 & -139.4 & 62.8 & 0.14 & 2002-OCT-13\\
CB 4  & IRAS 00362+5234 & -112.4 & 89.8 & 0.05 & 2003-MAY-10\\
CB 6  & IRAS 00465+6028 & -113.6 & 88.6 & 0.06 & 2003-MAY-10\\
CB 7  & IRAS 01078+6409 & -107.8 & 107.9 & 0.06 & 2004-AUG-04\\
      & IRAS 01087+6404 & -107.8 & 107.9 & 0.08 & 2004-AUG-04\\
      &                 & -107.7 & 108.0 & 0.04 & 2005-APR-14\\
CB 8  & IRAS 01202+7406 & -105.7 & 110.0 & 0.05 & 2004-AUG-04\\
CB 11 & IRAS 01333+6448 & -109.8 & 105.9 & 0.07 & 2004-AUG-04\\
      & IRAS 01334+6442 & -109.8 & 106.0 & 0.05 & 2004-AUG-04\\
      & IRAS 01341+6447 & -109.7 & 106.0 & 0.06 & 2004-AUG-05\\
CB 12 & CS peak         & -112.5 & 89.7 & 0.09 & 2003-MAY-09\\
      &                 & -119.4 & 96.4 & 0.05 & 2004-JUL-09\\
      & IRAS 01354+6447 & -119.3 & 96.4 & 0.05 & 2004-AUG-20\\
CB 15 & IRAS 03521+5555 & -107.2 & 108.5 & 0.08 & 2004-AUG-19\\
      & IRAS 03523+5608 & -107.2 & 108.5 & 0.10 & 2004-AUG-19\\ 
      &                 & -107.1 & 108.7 & 0.07 & 2005-APR-14\\
      & IRAS 03535+5555 & -107.2 & 108.5 & 0.05 & 2004-AUG-20\\ 
CB 16 & IRAS 03592-5642 & -2716 & 2677 & 0.05 & 2002-APR-09\\
      &                 & -103.3 & 98.9 & 0.10 & 2002-NOV-15\\ 
CB 17 & IRAS 04005+5647 & -112.8 & 103.0 & 0.05 & 2004-JUL-27\\ 
      & NH$_3$ peak     & -105.8 & 96.4 & 0.05 & 2003-MAY-09\\ 
CB 19 & IRAS 04233+2529 & -101.5 & 114.2 & 0.04 & 2004-AUG-20\\ 
      & IRAS 04240+2535 & -101.5 & 114.2 & 0.04 & 2004-AUG-20\\ 
CB 22 & NH$_3$ peak     & -105.6 & 110.1 & 0.05 & 2004-AUG-20\\
CB 23 & NH$_3$ peak     & -105.8 & 109.9 & 0.05 & 2004-AUG-20\\
      & CB23-cm1       & -102.6 & 113.1 & 0.06 & 2004-JUL-15\\
      &                 & -102.1 & 113.7 & 0.05 & 2005-APR-14\\
      & CB23-cm2       & -102.5 & 113.2 & 0.05 & 2004-JUL-27\\ 
CB 26 & IRAS 04559+5200 & -2722 & 2671 & 0.07 & 2002-APR-09\\
CB 27 & IRAS 05013+3234 & -101.2 & 114.6 & 0.08 & 2004-SEP-04\\ 
CB 28 & IRAS 05036-0359 & -99.2 & 116.5 & 0.10 & 2004-SEP-08\\ 
      & IRAS 05037-0402 & -99.3 & 116.5 & 0.12 & 2004-SEP-08\\ 
      & CS peak         & -92.3 & 109.9 & 0.3 & 2002-NOV-17\\ 
       &                & -92.3 & 109.9 & 0.07 & 2003-MAY-10\\
      & IRAS 05038-0400 & -99.3 & 116.5 & 0.07 & 2004-SEP-08\\ 
      & NH$_3$ peak     & -92.3 & 109.9 & 0.4 & 2002-NOV-17\\
       &                & -92.3 & 109.9 & 0.11 & 2003-MAY-09 \\
CB 29 & IRAS 05190-0348 & -96.8 & 119.0 & 0.06 & 2004-OCT-05\\ 
      &                 & -96.7 & 119.1 & 0.04 & 2004-OCT-31\\
      & IRAS 05194-0346 & -96.8 & 119.0 & 0.06 & 2004-SEP-08\\ 
      & IRAS 05194-0343 &-96.7 & 119.0 & 0.06 & 2004-OCT-05\\ 
      &                 & -96.7 & 119.0 & 0.05 & 2004-OCT-31\\
      & CS peak         &-89.9 & 112.3 & 0.4 & 2002-NOV-17\\ 
      &                 &-89.9 & 112.3 & 0.08 &  2003-MAY-10\\
      &                 & -96.6 & 119.1 & 0.04 & 2005-APR-14\\
      & IRAS 05201-0341 & -96.6 & 119.1 & 0.04 & 2004-OCT-31\\ 
CB 30 & [MYT99] CB 30 3 &-101.2 & 101.0 & 0.08 & 2003-MAY-11\\ 
      & IRAS 05274+0542 & -108.2 & 107.6 & 0.06 & 2004-OCT-05\\ 
      &                 &  -108.1 & 107.6 & 0.05 & 2004-OCT-30\\
      & IRAS 05268+0538 &-101.2 & 101.0 & 0.11 &  2003-MAY-08\\ 
      & IRAS 05268+0550 & -108.1 & 107.6 & 0.05 & 2004-OCT-05\\ 
      &                 &  -108.1 & 107.6 & 0.05 & 2004-OCT-29\\
CB 31 & [YMT96] CB 31 1 & -108.3 & 93.9 & 0.10 & 2003-MAY-08 \\ 
      & [YMT96] CB 31 2 & -108.3 & 93.9 & 0.08 & 2003-MAY-06 \\ 
      & IRAS 05307-0038 & -115.0 & 100.7 & 0.06 & 2004-OCT-05\\ 
      &                 & -115.0 & 100.8 & 0.07 & 2004-OCT-07\\
      &                 & -104.2 & 111.5 & 0.04 & 2004-OCT-26\\
CB 32 & IRAS 05344-0016 & -112.9 & 102.8 & 0.06 & 2004-OCT-31\\ 
      & CS peak         & -106.2 & 96.0 & 0.14 & 2003-MAY-09  \\ 
      &                 & -106.2 & 96.0 & 0.12 & 2003-MAY-10\\
CB 33 & IRAS 05433+2040 & -107.3 & 108.4 & 0.10 & 2004-SEP-04\\
      & IRAS 05437+2045 & -107.3 & 108.4 & 0.09 & 2004-SEP-04\\
CB 34 & [HSW99] CB 34 SMM 2 & -107.4 & 108.3 & 0.05 & 2004-AUG-20\\
      & [YMT96] CB 34 3 & -100.4 & 101.8 & 0.05 & 2003-MAY-06 \\
      & [HSW99] CB 34 SMM 1 & -93.9 & 121.8 & 0.06 & 2004-JUL-27\\
      & [HSW99] CB 34 SMM 5 & -107.4 & 108.3 & 0.06 & 2005-SEP-23\\
CB 37 & [MYT99] CB 37 1 &-99.9 & 102.3 & 0.10 & 2003-JUL-15 \\
      &                 & -107.0 & 108.7 & 0.07 & 2004-AUG-27\\
      &                 & -104.4 & 111.4 & 0.04 & 2005-APR-14\\
      & NH$_3$ peak     & -107.0 & 108.8 & 0.05 & 2004-AUG-20\\
      & [MYT99] CB 37 2 & -107.0 & 108.7 & 0.05 & 2004-AUG-20\\
CB 39 & IRAS 05591-1630 & -98.7 & 103.5 & 0.07 & 2003-MAY-10 \\
      & CS peak         & -98.7 & 103.5 & 0.14 &  2003-MAY-08\\
      & [YMT96] CB 39 2 & -98.7 & 103.5 & 0.05 & 2003-MAY-06\\
CB 40 & IRAS 05589+1638 & -118.9 & 96.9 & 0.07 & 2004-SEP-05\\
      & IRAS 05592+1640 & -105.4 & 110.3 & 0.10 & 2004-SEP-05\\
CB 42 & IRAS 06000+1644 & -105.2 & 110.6 & 0.08 & 2004-OCT-07\\
      &                 & -105.1 & 110.7 & 0.06 & 2004-OCT-26\\
CB 43 & [MYT99] CB 43 1 & -101.6 & 114.2 & 0.04 & 2005-OCT-16\\
CB 44 & CS peak         & -101.6 & 100.6 & 0.06 & 2003-MAY-07 \\
      & [MYT99] CB 44 2 & -101.6 & 100.6 & 0.06 & 2003-MAY-11\\
      & [MYT99] CB 44 3 & -101.6 & 100.6 & 0.07 & 2003-MAY-11 \\
      & IRAS 06047+1923 & -108.6 & 107.2 & 0.06 & 2004-OCT-07\\
      &                 & -108.3 & 107.4 & 0.10 & 2004-OCT-26\\
      & IRAS 06048+1934 &  -101.6 & 100.6 & 0.23 & 2003-MAY-08\\
CB 45 & IRAS 06055+1800 & -107.2 & 108.5 & 0.07 & 2004-OCT-07\\
      &                 & -107.2 & 108.5 & 0.05 & 2004-OCT-30\\
      & NH$_3$ peak     & -100.3 & 101.9 & 0.10 & 2003-MAY-07\\
CB 48 & IRAS 06175+0711 & -89.4 & 126.3 & 0.04 & 2004-OCT-30\\
CB 50 & IRAS 06316+0748 & -107.1 & 108.6 & 0.04 & 2004-OCT-30\\ 
CB 51 & IRAS 06355+0134 & -117.2 & 98.5 & 0.06 & 2004-DEC-01\\
CB 52 & [YMT96] CB 52 1 & -91.4 & 124.3 & 0.06 & 2004-OCT-30\\
      & IRAS 06464-1650 & -85.6 & 116.6 & 0.10 &  2003-MAY-07\\
      & IRAS 06464-1644 & -91.4 & 124.4 & 0.04 & 2005-FEB-19\\
      & IRAS 06471-1651 & -91.3 & 124.5 & 0.04 & 2005-FEB-19\\
CB 55 & IRAS 07019-1631 & -87.8 & 127.8 & 0.03 & 2005-FEB-19\\
CB 56 & IRAS 07125-2507 & -93.3 & 122.4 & 0.04 & 2005-FEB-19\\
      & IRAS 07125-2503 & -93.3 & 122.4 & 0.04 & 2005-FEB-19\\
CB 57 & IRAS 07154-2304 & -87.6 & 128.1 & 0.08 & 2005-FEB-22\\
      & IRAS 07156-2248 & -87.4 & 128.3 & 0.21 & 2004-DEC-01\\
      & IRAS 07156-2301 & -87.4 & 128.4 & 0.06 & 2005-FEB-19\\
CB 58 & IRAS 07159-2329 & -87.4 & 128.3 & 0.07 & 2005-FEB-22\\
      & [YMT96] CB 58 1 & -85.6 & 116.6 & 0.18 & 2003-MAY-04\\
      & IRAS 07161-2336 & -85.6 & 116.6 & 0.10 & 2003-MAY-07\\
CB 59 & IRAS 07171+0359 & -90.5 & 111.7 & 0.05 & 2002-APR-18\\
CB 60 & IRAS 08022-3115 & -93.9 & 121.8 & 0.07 & 2005-FEB-19\\
      & IRAS 08026-3122 & -94.0 & 121.8 & 0.23 & 2004-DEC-01\\
      & IRAS 08029-3118 & -93.0 & 121.8 & 0.11 & 2004-DEC-30\\
CB 63 & IRAS 15486-0350 & -2665 & 2728 & 0.02 & 2002-APR-09\\
CB 65 & [VRC2001] L1704 SMM 1& -132.6 & 83.1 & 0.09 & 2004-AUG-04\\ 
      & IRAS 16287-2337 & -98.8 & 103.4 & 0.16 & 2002-JUN-16\\ 
CB 67 & IRAS 16485-1906 & -130.3 & 85.4 & 0.08 & 2004-AUG-04\\ 
CB 68 & CS peak         & -102.6 & 113.1 & 0.08 & 2004-JUL-09\\ 
      &                 & -102.7 & 113.0 & 0.05 & 2005-JUN-06\\
      & IRAS 16544-1604 & -102.7 & 113.0 & 0.05 & 2004-JUL-08\\ 
      & NH$_3$ peak     & -102.8 & 113.0 & 0.10 & 2004-AUG-21\\ 
CB 78 & IRAS 17147-1821 & -2720 & 2673 & 0.06 & 2002-MAR-13 \\
CB 81 & outflow         & -104.3 & 111.4 & 0.15 & 2004-JUL-04\\
CB 82 &  [VRC2001] L57 SMM 1      & -104.2 & 111.5 & 0.3 & 2004-JUL-23\\ 
      &                 & -101.0 & 114.7 & 0.06 & 2005-JUN-06\\
      & NH$_3$ peak     & -104.4 & 111.3 & 0.12 & 2004-AUG-26\\ 
      & CS peak         & -97.5  & 104.7 & 0.06 & 2004-JUL-08\\ 
CB 98 & CB98-mm         & -97.0 & 118.7 & 0.14 & 2004-JUL-23\\ 
CB 100& IRAS 17490-0258 & -96.5 & 105.7 & 0.10 & 2002-NOV-16\\ 
      & IRAS 17493-0258 &-103.1 & 112.7 & 0.05 & 2004-AUG-21\\ 
CB 101& IRAS 17505-0828 & -2654 & 2739 & 0.02 & 2002-APR-10\\
      &                 & -101.1 & 114.6 & 0.07 & 2004-AUG-04\\ 
CB 104& IRAS 17526-0815 & -2654 & 2738 & 0.03 & 2002-APR-10\\
      & IRAS 17533-0808 & -97.1 & 118.6 & 0.05 & 2004-AUG-21\\
CB 105& IRAS 17561-0349 & -100.9 & 114.8 & 0.09 & 2004-AUG-21\\
CB 106& IRAS 17577-0329 & -91.5 & 124.2 & 0.06 & 2004-JUL-15\\
      & IRAS 17580-0330 & -101.5 & 114.2 & 0.05 & 2004-AUG-26\\
      & IRAS 17586-0330 & -101.5 & 114.3 & 0.06 & 2004-AUG-26\\
CB 108& [YMT96] CB 108 1& -95.5 &106.7 & 0.3 & 2002-SEP-30 \\
      &                 & -102.1 & 113.6 & 0.05 & 2005-JUN-06\\
CB 118& IRAS 18094-1550 & -2654 & 2739 & 0.03 & 2002-APR-08\\
CB 121& IRAS 18115-0701 & -101.9 & 113.9 & 0.08 & 2004-AUG-21\\
      & IRAS 18116-0657 & -101.9 & 113.8 & 0.09 & 2004-AUG-26\\
CB 124& IRAS 18120+0704 & -87.2 & 128.5 & 0.05 & 2004-JUL-15\\
      & IRAS 18122+0703 & -80.7 & 121.5 & 0.20 & 2002-NOV-15\\
CB 125 & NH$_3$ peak     & -94.7 & 107.5 & 0.8 & 2002-OCT-02\\
CB 128& IRAS 18132-0350 & -99.0 & 116.7 & 0.06 & 2004-JUL-07\\
CB 130& CB130-mm        & -97.6 & 118.1 & 0.06 & 2004-JUL-22\\
CB 131& CB131-smm       & -101.0 & 114.7 & 0.10 & 2004-JUL-24\\
CB 137& IRAS 18219-0100 & -98.4 & 117.3 & 0.10 & 2004-AUG-04\\
CB 142& IRAS 18272-1343 & -89.1 & 126.6 & 0.06 & 2004-JUL-08\\
CB 145& IRAS 18296-0911 & -103.1 & 112.7 & 0.05 & 2004-JUL-07\\
CB 146& IRAS 18293-0906 & -103.3 & 112.4 & 0.06 & 2004-JUL-09\\
      & IRAS 18294-0901 & -103.3 & 112.5 & 0.07 & 2004-JUL-09\\
      & [HSW99] CB 146 SMM 2& -103.3 & 112.4 & 0.05 & 2004-JUL-08\\ 
      & [HSW99] CB 146 SMM 1& -103.3 & 112.4 & 0.05 & 2004-JUL-08\\ 
CB 171& [YMT96] CB 171 3& -90.9 & 124.9 & 0.05 & 2004-JUL-22\\
      & [YMT96] CB 171 2& -90.9 & 124.8 & 0.06 & 2004-JUL-22\\
CB 177& IRAS 18599+1739 & -91.2 & 124.5 & 0.08 & 2004-SEP-04\\
CB 178& IRAS 18595+1812 & -92.2 & 123.5 & 0.13 & 2004-SEP-05\\
      & IRAS 19002+1755 & -92.2 & 123.5 & 0.07 & 2004-AUG-27\\
CB 180& NH$_3$ peak     & -95.9 & 119.9 & 0.04 & 2004-JUL-09\\
CB 184& IRAS 19116+1623 & -2655 & 2738 & 0.03 & 2002-APR-08\\
CB 187& IRAS 19162-0135 & -101.2 & 114.5 & 0.10 & 2004-SEP-03\\
CB 188& IRAS 19179+1129 & -99.9 & 115.8 & 0.07 & 2004-AUG-25\\
      & Outflow         & -100.4 & 115.3 & 0.07 & 2004-JUL-09\\
      & CS peak         & -94.0 & 108.2 & 0.10 & 2002-SEP-18\\
      & IRAS 19180+1127 & -100.4 & 115.3 & 0.05 & 2004-JUL-09\\
CB 189& IRAS 19180+1116 & -99.9 & 115.8 & 0.08 & 2004-AUG-25\\
      & [VRC2002] L673 SMM 7& -100.1 & 115.6 & 0.04 & 2005-JUN-06\\
      & [VRC2002] L673 SMM 1& -100.0 & 115.7 & 0.07 & 2005-FEB-26\\
      &                 & -100.8 & 115.0 & 0.05 & 2005-JUN-16\\
      & IRAS 19180+1114 & -100.0 & 115.8 & 0.05 & 2005-SEP-22\\
      & IRAS 19183+1123 & -100.4 & 115.3 & 0.06 & 2004-AUG-22\\
      & IRAS 19184+1118 & -100.5 & 115.3 & 0.06 & 2004-AUG-22\\
CB 190& IRAS 19186+2325 & -96.5 & 119.3 & 0.09 & 2004-AUG-22\\
CB 194& IRAS 19273+1433 & -103.7 & 112.1 & 0.12 & 2004-SEP-10\\
CB 196& IRAS 19329+1213 & -97.7 & 118.0 & 0.10 & 2004-SEP-09\\
CB 198& IRAS 19342+1213 & -97.7 & 118.0 & 0.06 & 2004-OCT-25\\
      &                 & -99.0 & 116.7 & 0.05 & 2005-JUN-16\\
CB 199& [ARC92] Barn 335 9& -99.2 & 116.5 & 0.09 & 2005-FEB-27\\
      & [ARC92] Barn 335 1& -108.9 & 106.8 & 0.05 & 2004-NOV-09\\
      & IRAS 19343+0727 & -99.2 & 116.5 & 0.08 & 2004-AUG-18\\
      & [ARC92] Barn 335 2& -108.9 & 106.8 & 0.06 & 2004-NOV-09\\
      &                   &  -99.2 & 116.5 & 0.10 & 2005-FEB-22\\
      & NH$_3$ peak     & -99.2 & 116.5 & 0.06 & 2004-AUG-18\\
      &                   & -99.2 & 116.5 & 0.12 & 2005-FEB-22\\
      &                   & -99.3 & 116.4 & 0.05 & 2005-JUN-16\\
      & IRAS 19345+0727 & -92.7 & 109.5 & 0.13 & 2003-MAY-16\\
      & [ARC92] Barn 335 11& -99.6 & 116.1 & 0.05 & 2005-JUN-15\\
      & IRAS 19347+0729 & -99.1 & 116.6 & 0.07 & 2004-SEP-03\\
      & IRAS 19348+0724 & -99.1 & 116.6 & 0.07 & 2004-SEP-04\\
CB 203& IRAS 19413+1902 & -91.7 & 124.1 & 0.09 & 2004-SEP-04\\
CB 205& IRAS 19427+2741 & -93.6 & 122.1 & 0.07 & 2004-AUG-22\\
      &                 & -85.1 & 117.2 & 0.06 & 2005-JUN-16\\
      & [YMT96] CB 205 1& -85.3 & 116.9 & 0.07 & 2002-SEP-03\\
      &                 & -85.3 & 116.9 & 0.10 & 2002-SEP-19\\
      &                 & -85.3 & 116.9 & 0.13 & 2002-NOV-16\\
      & [HSW99] L 810 SMM 2& -91.7 & 124.0 & 0.05 & 2004-JUL-22\\
      & [YMT96] CB 205 2& -2659 & 2734 & 0.03 & 2002-APR-08\\
      & IRAS 19433+2743 & -85.3 & 116.9 & 0.25 & 2003-JUL-11 \\
      &                 & -91.7 & 124.0 & 0.05 & 2004-JUL-22\\
      &                 & -91.8 & 124.0 & 0.05 & 2005-JUN-16\\
      & IRAS 19433+2751 & -91.7 & 124.1 & 0.07 & 2004-AUG-25\\
      & [YMT96] CB 205 3& -85.3 & 116.9 & 0.08 & 2002-SEP-03 \\
      &                 & -85.3 & 116.9 & 0.17  & 2002-NOV-16\\
      &                 & -91.7 & 124.0 & 0.05 & 2004-JUL-15\\
      & IRAS 19438+2757 & -91.6 & 124.1 & 0.07 & 2004-AUG-25\\
      & IRAS 19438+2737 & -91.7 & 124.0 & 0.10 & 2004-AUG-25\\
      & IRAS 19439+2748 & -91.7 & 124.1 & 0.07 & 2004-SEP-05\\
CB 206& [YMT96] CB 206 2& -92.2 & 123.5 & 0.07 & 2004-JUL-22\\
      &                 & -92.3 & 123.4 & 0.05 & 2004-AUG-05\\
CB 207& IRAS 19437+2108 & -99.9 & 115.9 & 0.06 & 2004-AUG-05\\
CB 208& IRAS 19450+1847 & -92.1 & 123.6 & 0.11 & 2004-SEP-10\\
CB 210& IRAS 19529+3341 & -97.8 & 118.0 & 0.09 & 2004-SEP-04\\
CB 211& IRAS 19576+2447 & -100.5 & 115.2 & 0.10 & 2004-SEP-04\\
CB 214& IRAS 20018+2629 & -91.2 & 111.0 & 0.14 & 2003-JUL-11\\
CB 216& Outflow         & -88.5 & 113.7 & 0.14 & 2002-SEP-17\\
      & IRAS 20037+2317 & -94.8 & 120.9 & 0.10 & 2004-SEP-05\\
      & CS peak         & -2659 & 2734 & 0.03 & 2002-APR-08\\
      &                 & -88.5 & 113.7 & 0.13 & 2002-SEP-17\\
CB 217& Outflow         & -100.6 & 101.6 & 0.12 & 2003-JUL-11\\
      & [YMT96] CB 217 4& -107.0 & 108.7 & 0.06 & 2004-JUL-24\\            
CB 219& IRAS 20176+6343 & -2680 & 2713 & 0.03 & 2002-APR-08\\
CB 222& IRAS 20328+6351 & -101.0 & 101.2 & 0.04 & 2002-APR-19\\
CB 224& IRAS 20355+6343 & -103.8 & 98.4 & 0.19 & 2003-JUN-07\\
      &                 & -110.3 & 105.3 & 0.04 & 2004-OCT-29\\
      &                 & -110.7 & 105.0 & 0.04 & 2005-JUN-05\\
      & NH$_3$ peak     & -2681 & 2713 & 0.04 & 2003-APR-08\\
CB 225& IRAS 20365+5607 & -103.6 & 98.6 & 0.06 & 2002-APR-18\\
      &                 & -103.6 & 98.6 & 0.13 & 2002-AUG-14\\
CB 230& [YMT96] CB 230 1& -98.2 & 104.0 & 0.04 & 2002-APR-19\\
      & [YMT96] CB 230 2& -2685 & 2708 & 0.05 & 2002-APR-07\\ 
CB 233& CS peak         & -108.5 & 107.3 & 0.10 & 2004-AUG-19\\
      &                 & -109.3 & 106.4 & 0.05 & 2005-APR-14\\
CB 235& IRAS 21548+5843 & -108.0 & 107.8 & 0.10 & 2004-AUG-19\\
CB 238& NH$_3$ peak     & -107.3 & 108.4 & 0.05 & 2004-AUG-22\\
CB 240& IRAS 22317+5816 & -111.5 & 104.3 & 0.08 & 2004-JUL-30\\
CB 241& IRAS 23095+6547 & -115.5 & 100.3 & 0.08 & 2004-JUL-30\\
CB 243& IRAS 23228+6320 & -112.2 & 90.0 & 0.09 & 2003-MAY-11\\
      & NH$_3$ peak     & -112.2 & 90.0 & 0.20 & 2002-OCT-13\\
      & [VRC2001] L1246 SMM 2& -119.0 & 96.8 & 0.07 & 2004-JUL-24\\
      & CS peak         & -119.0 & 96.8 & 0.06 & 2004-JUL-24\\
CB 244&[VRC2001] L1262 SMM 2& -104.0 & 111.8 & 0.08 & 2004-JUL-30\\
      & [YMT96] CB 244 1& -2694 & 2699 & 0.04 & 2002-APR-07\\
      & IRAS 23249+7406 &-97.2 & 105.0 & 0.12 & 2003-JUN-07\\
CB 246& CB246-mm        & -103.6 & 112.2 & 0.08 & 2004-JUL-24\\
CB 247& IRAS 23550+6430 & -104.8 &  97.4 & 0.06 & 2003-MAY-12\\

\enddata

\tablenotetext{a}{Velocity range covered by the observational
    bandwidth}
\tablenotetext{b}{Noise level, $1\sigma$}
\tablenotetext{c}{Date of observation}

\end{deluxetable}

\end{document}